\documentclass{article}
\pdfoutput=1

\date{March 18, 2019}

\usepackage[hidelinks]{hyperref}
\usepackage{proof,graphicx,stackrel,listings}
\makeatletter
\def\lst@PlaceNumber{\llap{\normalfont
                \lst@numberstyle{\the\lst@lineno}\kern\lst@numbersep}}
\makeatother

\newtheorem{theorem}{Theorem}
\newtheorem{proposition}[theorem]{Proposition}
\newtheorem{corollary}[theorem]{Corollary}
\newtheorem{definition}[theorem]{Definition}
\newtheorem{figdefn}{Definition}

\newcommand{\AGSquote}[3]{{\citep{Ptashne:AGS3}:#2:#3#1}}
\newcommand{\AGSshort}{monograph}
\newcommand{\AGSShort}{Monograph}
\newcommand{\AGSnoun}{the \AGSshort{}}
\newcommand{\AGSNoun}{The \AGSshort{}}

\newcommand{\constructive}{any proof of a compound property can be decomposed to proofs of constituent properties}

\newcommand{\citep}{\cite}

\title{Proofs of life: molecular-biology reasoning simulates cell behaviors from first principles}
\author{Ren\'e Vestergaard\thanks{Corresponding author: \texttt{renevestergaard@acm.org}}\\ (no current affiliation)\thanks{Work done in part at Research Center for Integrated Science, JAIST, Japan.} \and Emmanuel Pietriga\\ (Inria, CNRS \& Univ.\ Paris-Sud)}

\newcommand{\refSectTheory}{S1}
\newcommand{\refSectFormalReasoning}{S1.1}
\newcommand{\refSectIndDefn}{S1.1.1}
\newcommand{\refSectMetaconsist}{S1.1.4}
\newcommand{\refSectProveTruth}{S1.1.6}
\newcommand{\refSectDNF}{S1.1.7}
\newcommand{\refSectBigpict}{S1.1.9}
\newcommand{\refSectMachinery}{S1.3}
\newcommand{\refSectFormFct}{S1.3.1}
\newcommand{\refSectCoext}{S1.3.2}
\newcommand{\refSectCCD}{S1.4.1}
\newcommand{\refSectCCP}{S1.4.2}
\newcommand{\refSectPerturb}{S1.4.4}
\newcommand{\refSectInter}{S1.4.5}
\newcommand{\refSectSequential}{S1.5}
\newcommand{\refSectDiagrams}{S1.6}
\newcommand{\refSectElementary}{S1.6.1}
\newcommand{\refSectCoextSS}{S1.6.2}
\newcommand{\refSctSynchroDgrm}{S1.6.3}
\newcommand{\refSectPractice}{S2}
\newcommand{\refSectGenotype}{S2.1}
\newcommand{\refSectMIGRI}{S2.1.1}
\newcommand{\refSectSpecials}{S2.1.9}
\newcommand{\refSectPhysiology}{S2.2}
\newcommand{\refSectLambdaTrans}{S2.2.5}
\newcommand{\refSectPhenotype}{S2.3}
\newcommand{\refSectReasoning}{S2.3.2}
\newcommand{\refSectAGSsynchro}{S2.4.3}
\newcommand{\refSectCombinator}{S2.5}
\newcommand{\refSectExamples}{S2.5.1}
\newcommand{\refSectAbstracting}{S2.5.2}
\newcommand{\refSectRapid}{S2.5.3}
\newcommand{\refSectCIVariants}{S2.5.4}
\newcommand{\refSectCountercorrect}{S2.5.5}
\newcommand{\refSectAntiimmunity}{S2.5.7}
\newcommand{\refAppMotivation}{SA}
\newcommand{\refAppAxioms}{SB}
\newcommand{\refAppPremises}{SC}
\newcommand{\refAppPop}{SD}
\newcommand{\refAppCArg}{SE}
\newcommand{\refAppPArg}{SF}
\newcommand{\refAppObs}{SG}
\newcommand{\refAppSups}{SH}

\begin{document}

\maketitle
\begin{abstract}
We axiomatize the molecular-biology reasoning style, show compliance of the standard reference: Ptashne, \emph{A Genetic Switch}, and present proof-theory-induced technologies to help infer phenotypes and to predict life cycles from genotypes. The key is to note that `reductionist discipline' entails \emph{constructive} reasoning: \constructive{}. Proof theory makes explicit the inner structure of the axiomatized reasoning style and allows the permissible dynamics to be presented as a mode of computation that can be executed and analyzed. Constructivity and execution guarantee simulation when working over domain-spe\-cif\-ic languages. Here, we exhibit phenotype properties for genotype reasons: a molecular-biology argument is an open-system concurrent computation that results in compartment changes and is performed among processes of physiology change as determined from the molecular programming of given DNA. Life cycles are the possible sequentializations of the processes. A main implication of our construction is that formal correctness provides a complementary perspective on science that is as fundamental there as for pure mathematics. The bulk of the presented work has been verified formally correct by computer.
\end{abstract}

\section*{Introduction}

This article is antedisciplinary \cite{Eddy:Antedisciplinary2005}: it establishes formal scientific reasoning as a theoretical, as a practical, and as a hybrid pursuit, exemplified with molecular biology and in the same sense as for mathematics but with domain-specific implications. The article is also multi-disciplinary and integrative. The main contribution is the integration, which is expressed in a series of mathematical properties. We discuss the integration and its consequences from several perspectives. The text is written for the broadest possible scientific audience.

\subsection*{Big Picture: correctness as end vs as means}

Science is experiencing difficulty with its main correctness notion: statistical analysis that is intended to ensure reproducibility \citep{Ioannidis:PLoS2005}. The causes are partly inherent and partly systemic \citep{Ioannidis:PLoS2014}. The computer-verifiable formal correctness notion of mathematics is widely thought to be at cross purposes with science: 
\begin{quote}
\emph{``As far as the laws of mathematics refer to reality, they are not cer\-tain; and as far as they are certain, they do not refer to reality.''}~\citep{Einstein:GeoExp1921}
\end{quote}
Einstein \cite{Einstein:GeoExp1921} continues: 
\begin{quote}
\emph{``It seems to me that complete clearness as to this state of things first became common property through that new departure in mathematics which is known by the name of mathematical logic or ``Axiomatics.'' The progress achieved by axiomatics consists in its having neatly separated the logical-formal from its objective or intuitive content;''}~\citep{Einstein:GeoExp1921}
\end{quote}
We show that the logical-formal of molecular biology is predictive of the objective content. Specifically, that \textbf{the mathematification of the reasoning style of a reductionist scientific discipline is a first-principle theory of the subject matter}, cf.\ \citep{US-NRC:TheoryBiology}: molecular biology accounts for biology in terms of molecular interactions --- it is that simple. Reductionism guarantees that the reasoning structure matches the perceived subject-matter structure. With this, correctness means structure compliance, including undeviating use of premises (here: molecular programming of genomes), i.e., simulation, and several \emph{meta-theoretic} properties that are guaranteed for all (compliant) proofs, including technical aspects of first-\-prin\-ci\-ple\-ness. The two correctness notions are distinct but not independent. Indeed, formal correctness helps unlock open problems in biology: once things are technical, issues can be pursued paradigmatically.

\subsection*{Specifics: the role of structure}

``A satisfactory description or computation of [modular organization, complex ensemble behavior that might be called emergent behavior, and robustness in biological processes] is a critical challenge for the future of biology'' \citep{US-NRC:TheoryBiology}. Concretely, the cited report calls for ``effective conceptual treatment,'' ``a theory of constructive engineering principles of life,'' and ``a theoretical basis for how biological entities generate aggregates of higher complexity.'' By invoking standard logical meta-theory and adaptations of known computation theory, we show that molecular-biology reasoning answers the triple call, i.e., we complement the case for applied mathematics in natural science \citep{Wigner:UnreasonableEffectiveness60}: also pure-mathematics methodology can be unreasonably effective. Our approach is adapted from a century-long line of work on mathematics, including axiomatic reasoning \citep{PrincipiaMathematica}, the Curry-Howard Correspondence \citep{Howard:CurryHoward,Wadler:PropTypes2015}, and constructivity \citep{Bauer:AcceptingConst2017}. The line of work is one of several mathematical disciplines that rely on computers \citep{AvigadHarrison:CACM14,Hales:MathsTuring14}. \\

Towards the end of the 19th century, mathematicians started getting concerned about the use of increasingly advanced proof methods and the appearance of increasingly complex proofs. After several decades of foundational efforts on several fronts, Whitehead and Russell \cite{PrincipiaMathematica} succeeded in showing that a substantial body of mathematics could be undertaken using a small set of reasoning rules that seemingly avoided \emph{inconsistency} problems: rule clashes that allow everything to be inferred. The response was a change in attitude throughout pure mathematics so readers wanted to be convinced that proofs could be spelled out in axiomatic detail rather than be persuaded in an intuitive or a political sense \citep{Quinn:RevoMath12}. Accompanied by precise definitions that the reasoning could proceed from, one effect was to democratize authority, including for young researchers. While opaque to non-users, having step-by-step correctness available cuts down on the issues to consider at any given time and turns development into a game to be explored: ``[the methods] give valuable results in regions, such as infinite number, which had formerly been regarded as inaccessible to human knowledge'' \citep{PrincipiaMathematica}. To make the activity feasible, to scale to bigger proofs, to automate relevant parts, to not overlook details, and to ensure general high levels of confidence, formal proofs can be constructed and checked for rule compliance on a computer \citep{AvigadHarrison:CACM14,Hales_etal:Kepler2017}. The fully-axiomatic style of working affects what intermediate results get used and helps ``uncover new and rather elegant nuggets of mathematics'' via low-to-high level optimizations \citep{Gonthier:4CT08}. Modes of computation that guarantee formal correctness of its results are often made available, as \emph{proof by reflection} \citep{Gonthier:4CT08,Hales_etal:Kepler2017}, including the innate Curry-Howard, or `reasoning-as-computation' \citep{Howard:CurryHoward,Wadler:PropTypes2015}.

Constructivity (and its canonical form: intuitionism) is possibly the most subversive idea in mathematics \citep{Bauer:AcceptingConst2017}: it avoids appeals to what ought to/might as well be `true' for the stricter requirement that detailed justification is proffered: \constructive{}. For mathematics, it spurns the equivalent axioms of $p\vee\neg p$ (\emph{excluded middle}) and $\neg\neg p\rightarrow p$ (\emph{double-negation elimination}), for any formula, $p$, see Section~\refSectFormalReasoning{}. In the former case, neither $p$ nor $\neg p$ need be the case. In the latter, we would admit as justified anything additional ($p$) as long as this does not violate consistency ($\neg\neg p$). Our interest in constructivity is purely practical.

\subsection*{\emph{A \emph{reductionist} Genetic Switch}}

Molecular biology is the bottom-up view on gene regulation:
\begin{quote}
\emph{``Our goal is to understand gene regulation in terms of the interaction of molecules.''}~\AGSquote{/App.\refAppMotivation{}}{xiv}{23}
\end{quote}
Genes are the parts of an organism's DNA that are used to produce sequence-de\-ter\-mined molecules with a functional role for the organism. The functional form may be RNA or, later, \emph{protein}: polypeptides in specific conformation \cite{Crick:Nature1970}.
\begin{quote}
\emph{``Briefly put, the issue is as follows: all cells of a given individual organism inherit the same set of blueprints in the form of DNA molecules. But as a higher organism develops from a fertilized egg a striking variety of different cell types emerges. Underlying the process of development is the selective use of genes, the phenomenon we call gene regulation. At various stages, depending in part on environmental signals, cells choose to use one or another set of genes, and thereby to proceed along one or another developmental pathway. What molecular mechanisms determine these choices? The lambda life cycle is a paradigm for this problem: the virus chooses one or another mode of growth depending upon extracellular signals, and we understand in considerable detail the molecular interactions that mediate these processes. We believe that analogous interactions are likely to underlie many developmental processes; by establishing a description for the particular case of lambda, we develop ideas that inform other studies even though no other case looks exactly like lambda.''}~\AGSquote{/App.\refAppMotivation{}}{xiii}{11}
\end{quote}
The key DNA segment for the organism here looks as follows, see Section~\refSectGenotype{}:
\begin{center}
\includegraphics[width=.6\textwidth]{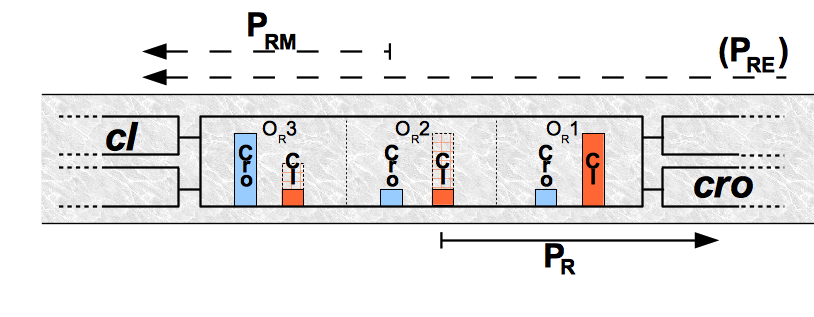}
\end{center}
The illustration shows the beginning of two genes (\emph{cI}, \emph{cro}), located on either side of a central region. The two genes are on separate DNA strands, which means they are transcribed in opposite directions: away from the central region. Transcription is done by \emph{RNA polymerase} (RNAP): a protein complex that has the ability to separate the two DNA strands, read off the particular base pairs on one of them, and assemble a strand of RNA that matches the base pairs in a particular way. DNA segments that allow RNAP to bind in the requisite manner are called \emph{promoters}. The illustration includes two promoters in the central region (P\textsubscript{RM}, P\textsubscript{R}) that permit the discussed transcription. The \emph{cI} gene may also be transcribed from a promoter that is located some distance away from the central region (P\textsubscript{RE}). The produced RNA may later be translated to polypeptides that may fold to proteins, in this case CI (aka repressor) and Cro. The central region also permits specific binding of the particular proteins, with the locations referred to as \emph{operators} (O\textsubscript{R}1, O\textsubscript{R}2, O\textsubscript{R}3). These bindings happen with different affinities, meaning at different effective concentrations, which serves as a basic determinant of the regulation of the two genes. The bindings may take place either intrinsically or cooperatively between operators, which affects both their affinities and durations, with varying effects on other bindings. For example, some operator bindings may prevent other bindings but they may also recruit RNAP to promoters that otherwise would rarely get occupied (dashed promoter lines). Some of the cooperative bindings involve operators that are separated by several thousand base pairs, which entails super-coiling of the DNA. We must be able to accurately account for all these specifics, see later. Presence of CI will be associated with the considered organism surviving by its DNA being embedded (\emph{lysogenized}) in and passively replicated alongside a host organism (\emph{the lysogenic cycle}). Presence of Cro is associated with the host's DNA-replication machinery being commandeered to produce on the order of a hundred copies of the parasite DNA, followed by host destruction (\emph{lysis}) and the release of viable parasite offspring (\emph{the lytic attack}). Interactions with the environment mean that the lytic attack results if many non-infected neighbouring hosts exist at the time of initial infection. Otherwise, the lysogenic cycle is established and maintained in the face of host replication, super-infection, and more. If an infected host has its DNA damaged, e.g., by UV light, the parasite has the ability to switch its mode of growth from an ongoing lysogenic cycle to the lytic attack (\emph{induction}). The combination of a (stable) lysogenic cycle and a (destructive) lytic attack with induction is called \emph{the temperate life cycle}.

\subsection*{Contents}

This article consists of a main body of text, end matter, and separate supplementary text, with its own sections and appendices. The main body is organized around our overall construction as applied to molecular biology. The key specifics for \AGSnoun{} are presented in figures that come with extensive legends. The figure legends are listed between the main body and end matter.

The article is accompanied by an instructional video: ``Using the CEqEA tool: \emph{A \emph{synthetic} Genetic Switch}''. The video is available at:\begin{center}\texttt{http://ceqea.sourceforge.net/extras/instructionalPoL.mp4}\end{center}

All technology we present, including the small \emph{kernel} to verify rule compliance of axiomatic molecular-biology reasoning, is available as free software: CEqEA, or the Cascaded-Equilibria Emergence Assistant \citep{CEqEA}. The requisite mathematical background is presented in Section~\refSectTheory{}, including computer verification of the key meta-theory of the tool by means of the Coq Proof Assistant \citep{coq}, see Section~\refSectMachinery{}. The meta-theory 1) guarantees that our axiomatization of molecular-biology reasoning makes sense, 2) clarifies fundamental molecular and structural aspects of biology as revealed in the reasoning, and 3) turns reasoning into engineering principles. Community-developed reasoning principles get selected for clarity, i.e., we ultimately rely on the power of simplicity \citep{Kelsey-al:R5RS}. 

We apply our molecular-biology technology to the standard reference: Ptashne \cite{Ptashne:AGS3}, referred to as `\AGSnoun{}', see above and Section~\refSectPractice{}. The reasoning in \AGSnoun{} is not as detailed as what we present, see Section~\refSectReasoning{}, but it is a close fit, see Appendices~\refAppMotivation{}--\refAppObs{}. In particular, it is only possible to make a factual distinction between our reasoning and \AGSnoun{} content in the case of statements that are superseded by additions to the latest edition \citep{Ptashne:AGS3}, see Appendix~\refAppSups{}. Irrespective of any particulars of our framework, the closeness means that \AGSnoun{} is coherent to an extent that rivals any mathematical text \citep{AvigadHarrison:CACM14}. Of the several hundred molecular-biology texts we have studied for this work, none appeared to match \AGSnoun{}'s level of precision. Our verification of \AGSnoun{} proceeded at the same rate as for similar mathematical texts: 1--2 days per page, excluding framework development but including time to deeply understand the text. It is not necessary to trust our tool's verification kernel: the proof details are listed in Section~\refSectReasoning{} and can be checked by hand.

\section{Overview}

We have identified 481 factual statements in \AGSnoun{} and classified them into eleven levels of abstraction, see Appendices~\refAppMotivation{}--\refAppSups{}. Five levels with 180 statements make up the organism-specific reasoning, see Theorem~\ref{thm:AvGS}. Two levels with 23 factual statements address general aims and means of molecular-biology reasoning, i.e., its foundation. One level covers counterfactual reasoning about select mutants: we address mutations systematically, with a corresponding focus. Two levels concern, e.g., DNA conformation and consequences: these statements provide background, remain implicit to the reasoning, and are not discussed here. The final level contains 18 reasoning statements that are implied to be superseded by statements added to the latest \AGSshort{} edition \citep{Ptashne:AGS3}.
\addtocounter{theorem}{-1}
\begin{theorem}[Overview]
\label{thm:AvGS}
Of the 180 reasoning statements in \AGSnoun{}, 81 (`population inferences', `causation arguments', `pathway arguments', `cell behaviors') follow progressively and axiomatically from the 99 (`molecular basis') within a logic that is consistent and compatible with the 23 foundation statements. The 18 superseded statements would require assumptions contradicting the 99 and are contradicted by inferences not covered by the 81.
\end{theorem}
The work extracts theory from practice, applies it, and develops meta-theory as a means to solve open problems, etc., see National Research Council (US) \citep{US-NRC:TheoryBiology}:
\addtocounter{theorem}{-1}\begin{theorem}[\ldots{}]
Biological reasoning over a `molecular basis' can be defined as open-system concurrent computation over the discrete manifestations of the molecule-coded regulation seen as `elementary processes of physiology change'. The mode of computation is induced from the reasoning rules' meta-theory-determined structural proof theory and \emph{simulates} reasoned-about cell behaviors: \textbf{given execution will exhibit the considered phenotype property for the specific reasons from the genotype}. Computation allows us to develop a phenotype for \AGSnoun{} or\-gan\-ism that incorporates the recent discoveries, is organized according to the coded-for dynamics, and revises but subsumes \AGSnoun{}'s observation-based account. Our prediction of the pos\-sible sequentializations of \AGSnoun{} concurrency reifies its life-cycle illustration.
\end{theorem}

\section{Principles}
In order to arrive at the axiomatization, we noted that the organism-specific reasoning in \AGSnoun{}, per the style of molecular biology, covers five progressive levels of abstraction from mol\-e\-cules to biology, see Theorem~\ref{thm:AvGS}. We:
\begin{enumerate}
\item\label{item:identify} identified the levels of abstraction used,
\item represented each level in a symbolic form, and
\item\label{item:connect} connected the symbolic forms while ensuring smoothness by:
\begin{enumerate}
\item\label{item:pure} fine-tuning the symbolic forms to be conceptually pure and distinct,
\item\label{item:abstract} abstracting each symbolic form to have a straightforward appearance,
\item\label{item:intermediate} introducing intermediate symbolic forms as necessary, and
\item\label{item:modality} parameterizing technical choices, e.g., as \emph{modalities}, see Section~\refSectGenotype{}.
\end{enumerate}
\end{enumerate}
Applying \ref{item:identify}--\ref{item:connect} is an exercise in taming concepts and combinatorics, which is a staple of the computer-science discipline of formal methods and of structured programming. Here, it was guided by the bottom-up nature of molecular biology, with existing staging already doing some constraining of technical issues. What we want to accomplish with \ref{item:identify}--\ref{item:connect} will have been increasingly accomplished with the maturation of a discipline. Pervasive abstraction, \ref{item:abstract}, facilitated smoothness and, with \ref{item:modality}, was necessary to account for all details in \AGSnoun{}, including interaction variations, cf.\ Guet et al.\ \cite{Guet:combinatorial2002}. All perceived non-linearity was seen to be the result of nested simplicity, see Wolpert \cite{Wolpert:unnatural1992}. Tautness of the permitted reasoning (see Theorem~\ref{thm:abintra}, Proposition~\ref{prop:lifecycle}) was helped by conceptual purity, \ref{item:pure}. The final axiomatization proceeds across a dozen stages, \ref{item:intermediate}, with the load distributed evenly in small and surprisingly-standard technicalities. 

\begin{figure}\scriptsize
\noindent (A)
\begin{verbatim}
+ !         |  $proteinC / cro   |  @PR_tr  ;                 //[15:13]
\end{verbatim}

\vspace*{.3cm}\noindent (B)
\begin{verbatim}
@PR_tr   =  RNAP.tr                                           //[14:6][6:-9a]
             |-- @OR1xor2_Cro                                 //[94:1][93:F4.24:1]
               : @OR1xor2_Cro +- !                            //[94:1][93:F4.24:1]
               : @OR1and2_Cro -- !                            //[94:1][93:F4.24:1]
               : @OLR12?_CI_rng  ! {|incdntl(CI)<mdtr,sink>}  //[111:F5.3:1][24:-8]
               ;
\end{verbatim}

\vspace*{.3cm}\noindent (C)
\begin{verbatim}
@OR1and2_Cro<@OR1_Cro,@OR2_Cro>  =  max&(@OR1_Cro,@OR2_Cro) ; //[93:F4.24:1]
@OR1xor2_Cro<> = min:(@OR1_Cro,@OR2_Cro) excl @OR1and2_Cro-- ;//[93:F4.24:1]
\end{verbatim}

\vspace*{.3cm}\noindent (D)
\begin{verbatim}
@OR1_Cro          =  @OR1_Cro_int     |--   @OLR12?_CI_rng ;  //intrinsic[17:6]
@OR2_Cro          =  @OR2_Cro_int     |--   @OLR12?_CI_rng ;  //intrinsic[17:6]
\end{verbatim}

\vspace*{.3cm}\noindent (E)
\begin{verbatim}
@OLR12?_CI_rng<>  =  @OLR123_CI--     : @OLR12_CI-- ;         //abbreviation
\end{verbatim}

\vspace*{.3cm}\noindent (F)
\begin{verbatim}
@OLR123_CI        =  (+) @OR3_CI_int  ;                       //[112:9]
@OLR12_CI         =  (+) @OR1_CI_int  excl  @OLR123_CI-- ;    //[86:F4.18][110:-4]
\end{verbatim}

\vspace*{.3cm}\noindent (G)
\begin{verbatim}
@OR1_CI_int<>     =  CI.l ;                                   //[86:F4.18:1]
@OR3_CI_int<>     =  CI.h ;                                   //[86:F4.18:1]
\end{verbatim}

\vspace*{.3cm}\noindent (H)
\begin{verbatim}
@OR1_Cro_int<>    =  Cro.h ;                                  //[25:F1.23:1]
@OR2_Cro_int<>    =  Cro.h ;                                  //[25:F1.23:1]
\end{verbatim}

\vspace*{.3cm}\noindent (I)
\begin{verbatim}
 $genomeI / cI    .  $proteinC / [ 0 {l} {m} {h} ] ;          //[12:-8]
 $genomeI / cro   .  $proteinC / [ 0 {l} {h} ] ;              //[12:-8]
\end{verbatim}
\caption{MIG/RI-specification: regulation of gene \emph{cro} in the $\lambda$\textsuperscript{[AGS3]} genotype\label{fig:specification}}
\end{figure}
For using the axiomatization, the starting point: how to formally represent the real world, requires considerable attention, as outlined in Figure~\ref{fig:specification}. Section~\refSectGenotype{} provides the full details, following best practices in formal methods: we prefix a formal language (MIG, for Modal Influence Graph) designed for transparency with a modeling methodology (RI, for Regulation Interface, see Section~\refSectMIGRI{}) designed for safety. We refer to the combination as MIG/RI specification. RI is given with motifs for capturing common types of molecular interaction, amounts to incrementally writing down organisms' molecular programming with no sec\-ond-guessing allowed, and involves working over states that are \emph{physiologically} determined: the nominal concentrations for which bindings may have an effect. Covariant, neutral, and contravariant effects are possible. Covariant effects derive from what is ostensibly specified, typically expression. Contravariant effects result from syntax annotations that our tool uses to determine, e.g., when decay explicitly outpaces production. Neutral effects may result both from lack of action and, e.g., from balanced decay and production. An example scenario: \AGSnoun{} organism contains a gene (\emph{cI}) that is transcribed from two promoters that require physiological presence in one of several configurations of specific recruiters of the transcribing molecules to have material effects. The promoters may be inhibited in several ways with varying rates of effect-from-inhibitor-concentration change. Additionally, the gene product may be subject to proteolysis by a host protein. The requisite notion of \emph{compound contravariance} is a combinatorial problem for MIG/RI specifications, cf.\ Ptashne \cite{Ptashne:NatChemBio11}, see Hay et al.\ \cite{HayHiggs:Superenhancer2016}. Relatedly, certain combinations of interactions may preclude each other. These are identified as logical contradictions and similar incompatibilities. Coded-for \emph{observable} concentrations are automatically inferred later, once everything that may affect the matter is known, see Proposition~\ref{prop:lifecycle}, cf.\ Guet et al.\ \cite{Guet:combinatorial2002}. Generally, a key objective at the starting stage is to avoid category mistakes: we use categories to pinpoint details and ensure components match their usages and are not conflated, cf.\ Ptashne \cite{Ptashne:PNAS13}.

\begin{figure}\scriptsize
\begin{figdefn}[Instance-sorted states, homogeneity] Consider some notion of \emph{states} sorted into \emph{schema instances}: different concentrations of a molecule population will, e.g., be states of the same sort. A set of states is \emph{homogeneous} if it does not contain two states of the same sort.
\end{figdefn}

\begin{figdefn}[Basics]$\ $
\begin{itemize}
\item $g$ ranges over formal genotypes, see Figure~\ref{fig:specification} and Section~\refSectGenotype{}.
\item $A$ and $R,M,P,I$ range over sets of states, with $A$ used for \emph{compartments}.
\item $c$, $R \stackrel[I]{M}{\rightarrow} P$ range over \emph{causations}: Reactants, Mediators, Products, Inhibitors, see Figure~\ref{fig:causations}.
\item $g\models R \stackrel[I]{M}{\rightarrow} P$ means `$g$ codes for $R \stackrel[I]{M}{\rightarrow} P$', see Figure~\ref{fig:causations} and Section~\refSectPhysiology{}.
\item \emph{Coinhibition} is causation-on-causation `inhibition', for when one of two opposing causations is stronger --- \emph{coinhibition-free($\_$)} is a predicate for its absence.
\item A genotype, $g$, is \emph{regular} if $g\models R \stackrel[I]{M}{\rightarrow} P$ implies that $R$ and $P$ contain one state each of the same instance sort, meaning if all causations are concentration changes, or similar.
\end{itemize}
\end{figdefn}

\begin{figdefn}[Positively-validated causations\label{defn:pos_cstn}]
$$
g\models^A R \stackrel[I]{M}{\rightarrow} P 
\quad\doteq\quad 
	(g\models R \stackrel[I]{M}{\rightarrow} P)
	\;\wedge\;(R \subseteq A)
	\;\wedge\;(M \subseteq A)
$$
\end{figdefn}

\begin{figdefn}[Coextension logic\label{defn:coextension}] Compartment changes, $\hookrightarrow$, are justified, $\vdash^{\mathrm{co}}$, from genotypes, $g$, by:
\begin{center}
\begin{tabular}{l}
\infer[\mbox{(interference)\hspace*{.5cm}if
$\left\{\begin{tabular}{l}
	$\{R_i \stackrel[I_i]{M_i}{\rightarrow} P_i\} \;\subseteq\; \{c \,\mid\, g\models^A c\}$\\[5pt]
	$\wedge\;\forall i \,.\, (I_i \cap A = \emptyset)$\\
	$\wedge\;\mbox{coinhibition-free}(\{R_i \stackrel[I_i]{M_i}{\rightarrow} P_i\})$\\[5pt]
	$\wedge\;\mbox{homogeneous}((A \setminus \bigcup_i R_i)\cup\bigcup_i P_i)$
\end{tabular}\right.$}]
{g\vdash^{\mathrm{co}} A{\hookrightarrow}(A \setminus \bigcup_i R_i)\cup\bigcup_i P_i}{}
\\[10pt]
\infer[\mbox{(sequence)}]{g\vdash^{\mathrm{co}} A_a{\hookrightarrow}A_c}{g\vdash^{\mathrm{co}} A_a{\hookrightarrow}A_b & g\vdash^{\mathrm{co}} A_b{\hookrightarrow}A_c}
\end{tabular}
\end{center}
\end{figdefn}
\caption{Coextension logic (default modalities)\label{fig:logic}}
\end{figure}
\begin{figure}\scriptsize
\noindent (A)
\begin{center}\begin{tabular}{c@{\hspace{1cm}}c@{\hspace{1cm}}c}
${\mathtt{Cro.0} \stackrel[\mathtt{CI.\{l,m,h\}}]{\mathtt{RNAP.tr}}{\rightarrow} \mathtt{Cro.l}}$
& ${\mathtt{Cro.l} \stackrel[\mathtt{CI.\{l,m,h\}}]{\mathtt{RNAP.tr}}{\rightarrow} \mathtt{Cro.h}}$
& ${\mathtt{Cro.h} \stackrel[\mathtt{CI.\{l,m,h\}}]{}{\rightarrow} \mathtt{Cro.l}}$\\[10pt]
${\mathtt{Cro.h} \stackrel{\mathtt{CI.h}}{\rightarrow} \mathtt{Cro.l}}$
& ${\mathtt{Cro.h} \stackrel{\mathtt{CI.m}}{\rightarrow} \mathtt{Cro.l}}$
& ${\mathtt{Cro.h} \stackrel{\mathtt{CI.l}}{\rightarrow} \mathtt{Cro.l}}$\\[5pt]
${\mathtt{Cro.l} \stackrel{\mathtt{CI.h}}{\rightarrow} \mathtt{Cro.0}}$
& ${\mathtt{Cro.l} \stackrel{\mathtt{CI.m}}{\rightarrow} \mathtt{Cro.0}}$
& ${\mathtt{Cro.l} \stackrel{\mathtt{CI.l}}{\rightarrow} \mathtt{Cro.0}}$
\end{tabular}\end{center}

\vspace*{.3cm}\noindent (B)
\begin{quote}
\emph{``In a lysogen, repressor bound at O\textsubscript{R}1 and O\textsubscript{R}2 keeps cro off while it stimulates transcription of its own gene cI''}~\AGSquote{/App.\refAppCArg{}}{22}{1}
\end{quote}
\begin{quote}
\emph{``More Cro is made until it reaches a level at which O\textsubscript{R}1 and O\textsubscript{R}2 are also filled and polymerase is prevented from binding to P\textsubscript{R}''}~\AGSquote{/App.\refAppCArg{}}{25}{7}
\end{quote}

\vspace*{.3cm}\noindent (C)
\begin{eqnarray*}
\lceil\verb/e1 |-- e2/\rceil &\doteq& (\lceil\verb/e1/\rceil\vee\neg\neg\lceil\verb/e2/\rceil)\wedge\neg\lceil\verb/e2/\rceil\\
\mbox{2DNF}(\lceil\verb/o2.s |-- (o1.s |-- o0.s)/\rceil) &=& (o_2.s\wedge \neg o_1.s \wedge \neg\neg\neg o_0.s)\vee(o_2.s\wedge\neg\neg o_0.s)
\end{eqnarray*}

\vspace*{.3cm}\noindent (D)
\begin{eqnarray*}
\mathrm{NI}(0) &\doteq& o_0.s\\
\mathrm{NI}(n+1) &\doteq& o_{n+1}.s\ \mbox{\texttt{|-{}-}}\; (\mathrm{NI}(n))
\end{eqnarray*}

\vspace*{.3cm}\noindent (E)
\begin{eqnarray}\mbox{\#cstn}(\mathrm{NI}(n)) &=& \emph{F}_{n+1} \hspace{.95cm}(\mbox{where }F_1 \doteq 1; F_2 \doteq 1; F_{n+2} \doteq F_{n+1}+F_n)\nonumber\\
\nonumber\\
\mbox{\#cstn}(\mathrm{NI}(n+1)) 
&=& \mbox{\#cstn}((o_{n+1}.s \,:\, \neg\neg\mathrm{NI}(n)) \,\&\, \neg\mathrm{NI}(n))\label{line:dual1}\\
&=& \mbox{\#cstn}((o_{n+1}.s \,\&\,\neg\mathrm{NI}(n)) \;:\; (\neg\neg\mathrm{NI}(n)\,\&\,\neg\mathrm{NI}(n)))\label{line:dual2}\\
&=& \mbox{\#cstn}(o_{n+1}.s \,\&\,\neg\mathrm{NI}(n)) + \mbox{\#cstn}(\neg\neg\mathrm{NI}(n)\,\&\,\neg\mathrm{NI}(n))\label{line:dual3}\\
&=& \mbox{\#cstn}(\neg\mathrm{NI}(n)) + 0\label{line:dual4}\\
\nonumber\\
\mbox{\#cstn}(\neg\mathrm{NI}(n+1)) 
&=&  \mbox{\#cstn}((\neg o_{n+1}.s \,\&\, \neg\neg\neg\mathrm{NI}(n)) \,:\, \neg\neg\mathrm{NI}(n))\label{line:rr5}\\
&=&  \mbox{\#cstn}(\neg o_{n+1}.s \,\&\, \neg\neg\neg\mathrm{NI}(n)) +  \mbox{\#cstn}(\neg\neg\mathrm{NI}(n))\label{line:rr6}\\
&=&  \mbox{\#cstn}(\neg\neg\neg\mathrm{NI}(n)) +  \mbox{\#cstn}(\neg\neg\mathrm{NI}(n))\label{line:rr7}\\
&=&  \mbox{\#cstn}(\neg\mathrm{NI}(n)) +  \mbox{\#cstn}(\mathrm{NI}(n))\label{line:rr8}
\end{eqnarray}
\caption{Causations vs `causation arguments' for protein Cro; nested inhibition\label{fig:causations}}
\end{figure}
The specifics of the last two stages ---which is what a reasoner works with--- are due to us: \emph{causations} and \emph{com\-part\-ment changes}, see Figure~\ref{fig:logic}. Causations consolidate the information in `causation arguments', see Figure~\ref{fig:causations}, i.e., resulted from \ref{item:pure},\ref{item:abstract}, and include consideration of the difference between their construction (as the discrete manifestations of regulation, typically concentration changes) and their use (as elementary processes of physiology change, typically changes to the bindings that the gene product may be involved in). Causations have formative relevance to later technologies, see Propositions~\ref{prop:automated},\ref{prop:lifecycle}, and seemingly to biology itself, see Discussion. Compartment chan\-ges sit between `pathway arguments' and `cell behaviors', see Figure~\ref{fig:derivations}, i.e., resulted from \ref{item:pure},\ref{item:intermediate}.\\
\begin{figure}\scriptsize
\noindent (A)
\begin{center}\begin{tabular}{c}
$\{\,\mathtt{DNA.ss},\, \mathtt{RNAP.tr},\, \mathtt{RecA.},\, \mathtt{RecA.*},\, \mathtt{CI.0},\, \mathtt{CI.l},\, \mathtt{CI.m},\, \mathtt{CI.h},\, \mathtt{Cro.0},\, \mathtt{Cro.l},\, \mathtt{Cro.h}\,\}$
\end{tabular}\end{center}

\vspace*{.5cm}\noindent (B)
\begin{center}\scriptsize\begin{tabular}{c}\{\qquad
$[]$:${\mathtt{RecA.*} \stackrel[\mathtt{DNA.ss}]{}{\rightarrow} \mathtt{RecA.}}$ \quad;\quad
$\lbrack\mathtt{SOS}\rbrack$:${\mathtt{CI.l} \stackrel{\mathtt{RecA.*}}{\rightarrow} \mathtt{CI.0}}$ \quad;\quad
$\lbrack\mathtt{SOS}\rbrack$:${\mathtt{CI.0} \stackrel{\mathtt{RecA.*}}{\rightarrow} \mathtt{CI.0}}$ \quad;\\
$\lbrack\mathtt{PRM\_tr}\rbrack$:${\mathtt{CI.l} \stackrel[\mathtt{Cro.\{l,h\}}]{\mathtt{CI.l},\mathtt{RNAP.tr}}{\rightarrow} \mathtt{CI.m}}$ \quad;\quad
$\lbrack\mathtt{PR\_tr}\rbrack$:${\mathtt{Cro.0} \stackrel[\mathtt{CI.\{l,m,h\}}]{\mathtt{RNAP.tr}}{\rightarrow} \mathtt{Cro.l}}$
\qquad\}\end{tabular}\end{center}

\vspace*{.5cm}\noindent (C)
\begin{center}\begin{tabular}{c}
$[\mathtt{PRM\_tr}]\;\; \mbox{\texttt{|-{}-}}\;\; [\mathtt{SOS}]$
\end{tabular}\end{center}

\vspace*{.5cm}\noindent (D) 
\begin{center}\begin{tabular}{c}
\infer{g\vdash^{\mathrm{co}} \{\mathtt{CI.l,Cro.0,DNA.ss,RecA.*}\}{\hookrightarrow}\{\mathtt{CI.0,Cro.l,DNA.ss,RecA.*}\}}
	{
		\infer{g\vdash^{\mathrm{co}} \{\mathtt{CI.l,Cro.0,.ss,.*}\}{\hookrightarrow}\{\mathtt{CI.0,Cro.0,.ss,.*}\}}{
[\![\mathtt{[SOS]}\,;\, \mathtt{{[]},{[PR\_tr]},{[PRM\_tr]}}]\!]
		}
	&
		\infer{g\vdash^{\mathrm{co}} \{\mathtt{CI.0,Cro.0,.ss,.*}\}{\hookrightarrow}\{\mathtt{CI.0,Cro.l,.ss,.*}\}}{
[\![\mathtt{{[PR\_tr]},{[SOS]}}\,;\, \mathtt{[]}]\!]
		}
	}
\end{tabular}\end{center}

\vspace*{.5cm}\noindent (E) 
\begin{center}
\begin{minipage}[t]{0.56\textwidth}
\begin{lstlisting}[basicstyle=\scriptsize\ttfamily,title={\scriptsize Linearized coextension derivation Section~\refSectReasoning{}:2/c/},captionpos=b]
[CI.l, Cro.0, RecA.*] + [DNA.ss]
-> [SOS]
x> ![] |-- [DNA.ss] ;                [PR_tr] |-- [CI.l!@[OLR12,OLR123]] ; [OLR12_CI, PRM_tr] |-- [SOS]

[Cro.0, RecA.*] + [CI.0, DNA.ss]
-> [PR_tr] ; [SOS]
x> ![] |-- [DNA.ss]

[Cro.l, RecA.*] + [CI.0, DNA.ss]
\end{lstlisting}
\end{minipage}
\hspace*{0.02\textwidth}
\begin{minipage}[t]{0.34\textwidth}\vspace*{.05cm}\small
\emph{``Two changes result [from the SOS response's RecA*-mediated cleavage of CI]. First, as repressor vacates O$_R$1 and O$_R$2 the rate of repressor synthesis drops (because repressor is required to turn on transcription of its own gene); and second, polymerase binds to P$_R$ to begin transcription of cro.''}\\[3pt]
\scriptsize `Pathway arg.'\ \hspace*{0cm} \AGSquote{/App.\refAppPArg{}}{24}{-7}
\end{minipage}
\end{center}
\caption{Coextension derivations vs `pathway arguments'\label{fig:derivations}}
\end{figure}

For precise definitions, well-formedness is always a primary concern. In case of a reasoning axiomatization, we need to consider, e.g., inadvertent admittance of paradoxes. The problem is that simple rules do not necessarily result in simple behaviors when put together, or even in isolation. \emph{Consistency} (Theorem~\ref{thm:consistency}) guarantees that  a given system of rules makes distinctions: some form of sense is being made. Paradoxes would make everything provable, see Section~\refSectMetaconsist{}.

\begin{theorem}[Consistency\label{thm:consistency}] \emph{Coextension logic} (Figure~\ref{fig:logic}) cannot justify arbitrary compartment changes.
\end{theorem}
To address the sense being made, we note that molecular-biology reasoning is scientifically reductionist:
\begin{quote}
\emph{``Biologists work on systems that have evolved. This gives us hope that any given case can be understood reductively. Nature built the system in steps, each step making an improvement on the previous version and so, this line of thought goes, the investigator can take it apart, study it in bits, and, perhaps, see how it all fits together.''}~\AGSquote{/App.\refAppAxioms{}}{xi}{1}
\end{quote}
The corresponding meta-theoretic property for axiomatizations is \emph{constructivity}: \textbf{\constructive{}}. A key insight here is that constructivity affirms Ptashne's `perhaps': the justification for a biological property is always meticulously and exclusively stated in terms of molecular interactions. Decomposition cannot be guaranteed, e.g., with state-space modeling and truth-based reasoning, meaning properties there may have spurious justifications, see Discussion.

\begin{theorem}[Coding\label{thm:coding}] Assume states, $o.s$, in a genotype, $g$, are \emph{testable}:\linebreak ${\neg o.s\vee\neg\neg o.s}$ holds. If ${g\vdash^{\mathrm{co}} A_0\hookrightarrow A_n}$ can be justified in coextension logic then the propositional implication ${\lceil g\rceil\rightarrow\lceil A_0\rceil\rightarrow\lceil A_n\rceil}$ is provable in the canonical constructive logic: \emph{intuitionistic logic}, where $\lceil\_\rceil$ encodes coextension syntax as (varying) mathematical formulas. In particular, genotypes code for compartment changes in the computable sense, up to timing, stochasticity, and divergence.
\end{theorem}

The states in a MIG/RI-specification represent molecular ground notions, typically concentrations. Testability is an extra-logical \emph{open-system assumption} prescribing that populations are either considered at a different concentration than the one at hand ($\neg o.s$) or no conflict arises if we insist on the one ($\neg\neg o.s$): it is already there or the population was not being considered. Constructively, not-not does not cancel out, meaning $\neg\neg o.s$ and $o.s$ are not identical, see Discussion.

The stronger \emph{decidability} assumption for states: ${o.s\vee\neg o.s}$ holds, is invoked with state spaces and truth-based reasoning. Subverting testability, decidability is an extra-logical closed-system assumption, see Sections~\refSectProveTruth{},\refSectBigpict{}, cf.:
\begin{quote}
\emph{``We wish to understand which steps are controlled by internal cellular programs and which by extracellular signals.''}~\AGSquote{/App.\refAppAxioms{}}{3}{1}
\end{quote}

While the result establishes that molecular-biology reasoning is a special case of mathematical reasoning, we need more to avoid false-positive verification.

The Curry-Howard Correspondence (CHC) establishes that an intuitionistic proof is a computable function that does what is proved \citep{Howard:CurryHoward,SorensenUrzyczyn:CH2006,Wadler:PropTypes2015}. CHC is the mathematics version of the reasoning-computation correspondence for molecular biology we develop here. Theorem~\ref{thm:coding} shows that there exists a computable function that takes as arguments a genotype, $g$, two compartments, $A_0$,$A_n$, and evidence that the three are coextension related: ${g\vdash^{\mathrm{co}} A_0\hookrightarrow A_n}$. With the $g$-coded causations as building blocks, it then mimics the compartment change. The function could be presented as a terminating Turing Machine \citep{Wadler:PropTypes2015}, cf.\ Brenner \cite{Brenner:Turing2012}. Call it \textsc{MolBio}$_{\mathrm{TM}}$. It remains terminating but becomes organism-specific when given a $g$: \textsc{MolBio}$_{\mathrm{TM}}^g$, etc. This means that genotypes translate to computable behavior when we have a way of taking an $A_0$ and producing a derived $A_n$, see Proposition~\ref{prop:automated}. 1) The result constrains the sense that coextension logic makes to something reasonable: provided choices are fixed at the start, a genotype, $g$, can accomplish only what \textsc{MolBio}$_{\mathrm{TM}}^g$ can, based exclusively on information in $g$, hence `coding'. 2) The result does not address our ultimate concern: the means of computation, including anything pertaining to choices. The mathematics means are a form of \emph{closed-system sequential computation}, as seen with CHC. The molecular-biology means are a form of \emph{open-system concurrent computation}, as we show. `Open system' refers to the fact that molecular-biology reasoning explicitly considers choices that are not made up front but may come at any time, including from the environment. Specifically, `timing' concerns the relative durations of causations, `stochasticity' is the possibility that causations may proceed from concentrations other than the nominal physiological ones, and `divergence' refers to conflicting causations. We return to the issues in the relevant technical context following Proposition~\ref{prop:automated}.\\

The constructivity in Theorem~\ref{thm:coding} is indirect: its decomposition of proofs is relative to mathematical formulas. We also have direct constructivity.
\begin{proposition}[Constructivity\label{prop:constructive}] Any compartment change from a regular genotype, $g$, can be justified by interfering $g$-coded causations (in an arbitrary order).
\end{proposition}

\section{Engineering: inner reasoning structure}
Logical meta-theory is not mathematics-specific and applies here, too. As we shall discuss, the meta-theory actively helps us extract simulation machinery. With molecular biology being concrete, several issues assume special interest.

\begin{theorem}[Consequence\label{thm:consequence}] For regular genotypes, coextension logic is \emph{a logic}: anything assumed follows as a consequence and if all of one set of assumptions are consequences of another then all consequences of the first are also consequences of the second. In formal shorthand, with implicit outer `forall's ($\forall$):
\begin{center}
$\begin{array}{c}
(c\in g) \rightarrow (g\vdash^{\mathrm{co}} c)\\[5pt]
\mbox{and}\\[5pt]
(\forall c\in g_1 \,.\, g_2\vdash^{\mathrm{co}} c)\rightarrow (g_1\vdash^{\mathrm{co}} A_0\hookrightarrow A_n) \rightarrow(g_2\vdash^{\mathrm{co}} A_0\hookrightarrow A_n)
\end{array}$
\end{center}
\end{theorem}
In principle, Theorem~\ref{thm:consequence} is a basic sanity check akin to Theorem~\ref{thm:consistency}, but its justification includes its proof and corollaries. Given a reasoning axiomatization, `structural proof theory' refers to the inner structure of the axiomatized reasoning style. Our core point is that \textbf{inner reasoning structure constitutes `en\-gi\-neer\-ing principles' for the subject matter in case of (direct) constructivity}, see National Research Council (US) \citep{US-NRC:TheoryBiology}. A key methodological point is that the properties in Theorem~\ref{thm:consequence} help establish structural proof theory. For example, the property in Corollary~\ref{cor:cut} follows from Theorem~\ref{thm:consequence} but is often difficult to establish by itself, partly because it reveals that reasoning in a logic allows for manipulation of assumptions even if the specific rules do not, as here. 
\begin{corollary}[Cut\label{cor:cut}] For regular genotypes, the cut rule is \emph{admissible} for coextension logic: adding it to the defining rules would not allow us to justify more compartment changes. In formal shorthand, with \textvisiblespace$\oplus[c]$ meaning `the (non-depleted) genotype that also codes for $c$':
\[\infer[\mbox{(cut)}]
  {g\vdash^{\mathrm{co}} A_0\hookrightarrow A_n}
  {g\vdash^{\mathrm{co}} c
  & g\oplus[c]\vdash^{\mathrm{co}} A_0\hookrightarrow A_n}\]
(Cut is not used for any reasoning here nor is it added to Figure~\ref{fig:logic}. The formal result is in a more restricted form, see Section~\refSectCoext{}.)
\end{corollary}
Concretely, the cut rule addresses when the manifestations of genes may take the place of the genes themselves, e.g., medicinal intervention ($\ldots\vdash^{\mathrm{co}} c$) to ensure unchanged cytoplasm operation ($\ldots\vdash^{\mathrm{co}} A_0\hookrightarrow A_n$) in case of a depleted genotype ($g\vdash^{\mathrm{co}}\ldots$). This is not a simple issue. More work on cut remains.

\begin{figure}\scriptsize
\begin{center}
\includegraphics[width=\linewidth]{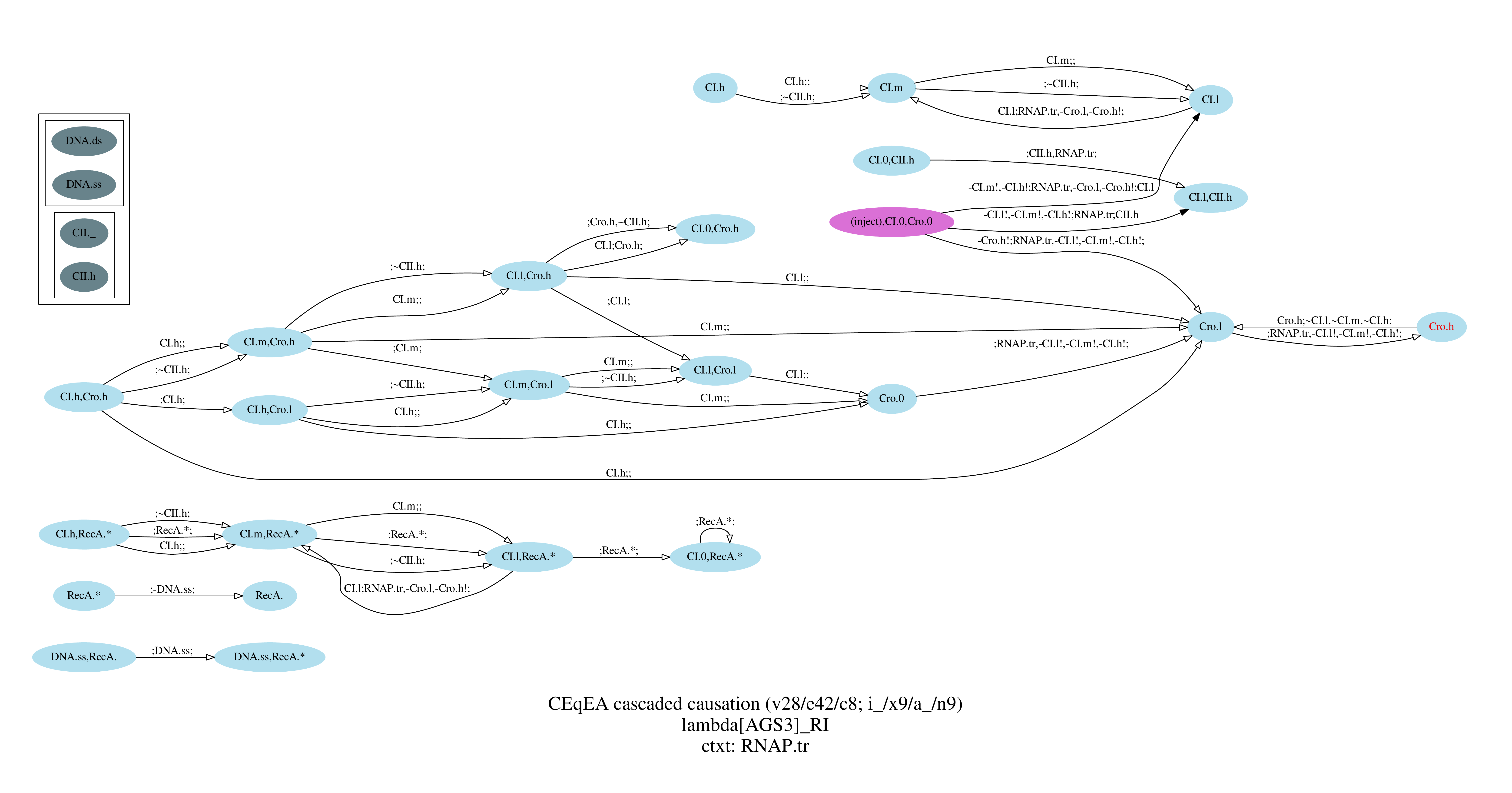}
\end{center}
\caption{$\lambda$\textsuperscript{[AGS3]} cascaded causation diagram\label{fig:diagram}}
\end{figure}

The properties in Theorem~\ref{thm:consequence} are an example of induction loading. When a result (Corollary~\ref{cor:cut}) is difficult to prove, a stronger result (Theorem~\ref{thm:consequence}) can be easier owing to stronger induction hypotheses: an analytic insight into the abstract nature of the problem is being expressed synthetically in the stronger result. Theorem~\ref{thm:consequence} captures that the key issue in structural proof theory, i.e., in understanding how proofs hang together, is the interplay between assumptions (here: genes) and their usage (here: expression), with the proof having to account for how the interplay propagates across the axiomatization (here: compartment dynamics). To address the issue, we defined (and proved equivalent, see Propositions~\ref{prop:constructive},\ref{prop:normal}) a version of coextension logic where interference is between either nothing, one causation, or two interferences, i.e., where interferences are arbitrarily ordered. Ordered interference enjoys smoother proofs of the two properties, i.e., ordering serves to separate concerns. The other concerns amount to order manipulation.
\begin{proposition}[Normalization\label{prop:normal}] Causation interferences for a compartment change that have been justified in a specific order can be justified concurrently.
\end{proposition}
Without Propositions~\ref{prop:constructive},\ref{prop:normal}, distributed/ordered vs.\ concurrent/un\-ordered interference is problematic: which is right? We use unordered interference because it is the more direct, has the least administrative overhead, see Figures~\ref{fig:logic},\ref{fig:derivations}, and allows us to reason over the graphs of causations: \emph{cascaded causation diagrams}, see Figure~\ref{fig:diagram}. Graphs are good for visualization, automation, and analysis.
\begin{proposition}[Automated reasoning\label{prop:automated}] Letting a cascaded causation diagram for a genotype, $g$, act successively on compartment $A_0$ to change it to $A_n$ builds a coextension derivation of $g\vdash^{\mathrm{co}} A_0\hookrightarrow A_n$, see Figure~\ref{fig:derivations} and video.
\end{proposition}
A cascaded causation diagram \emph{acts} on a compartment by considering the nodes whose content is in the compartment; considering their out-edges; excluding edges that are {(co-)}\-inhibited; and updating the compartment with the states from the targeted nodes, provided the result is homogeneous, see Section~\refSectCCP{}. For reasons we discuss along Proposition~\ref{prop:lifecycle}, we call the action \emph{CCP} (Calculus of Coextensive Processes). To be exact, it is CCP\textsuperscript{all}, where `all' is a strategy for choosing out-edges. The `all' strategy is appropriate when \emph{timing} is trivial: all causations proceed at the same timescale, and \emph{stochastic} behaviors are ignored: operators only function in line with the concentrations that nominally lead to their differential occupancy. Our CCP\textsuperscript{all} implementation is interactive in part to allow timing and stochastic issues to be modeled manually, as needed, see video. We say we have encountered a \emph{divergence} when a targeted compartment is not homogeneous, i.e., if two causations attempt to send some population towards different concentrations or a causation attempts to undermine a (non-auto-)mediating population. Coextension logic is designed to allow for reasoning about all that \emph{may} happen while cascaded causation diagrams and CCP\textsuperscript{all} provide a convenient platform for attempting to predict what \emph{will} happen. 

Apropos of Theorem~\ref{thm:coding}, we note that \AGSnoun{} organism exhibits only CCP\textsuperscript{all}-divergences involving mediation and only in a few edge cases that are easily resolved. As seen when we do rapid phenotyping below, mutants of \AGSnoun{} organism exhibit divergences proper as a result of overlapping cooperative bindings that are persistent and mutually exclusive. We would imagine programmatic/diver\-gence-free behavior (pathways, to a first approximation) is pervasive within evolved organisms, cf.\ \AGSquote{/App.\refAppAxioms{}}{3}{1} and Brenner \cite{Brenner:Turing2012}.

\section{Emergence: partitioning the reasoning space}
CCP\textsuperscript{all} halts when encountering a divergence, an absence of applicable causations, or if directed to do so by MIG/RI-specified expiration (here: to capture host lysis). We can also manually interrupt CCP\textsuperscript{all}, e.g., to break loops, and we can initiate or alter progress to address specific points of interest, e.g., fresh/super infection, (loop) perturbation, and environmental signals. Progress alteration may come at the price of a new coextension derivation, see Proposition~\ref{prop:automated}.
\begin{figure}\scriptsize
\begin{description}
\item[Host is healthy or not:] $\heartsuit$ stands for `absent a host SOS response', $\neg\heartsuit$ its negation `during a host SOS response';
\item[Host is doomed or not:] $\spadesuit$ stands for `with \texttt{CII} below high concentration and with \texttt{CI} (repressor) below physiological concentration', $\neg\spadesuit$ its negation `with \texttt{CII} at high concentration or with \texttt{CI} at physiological concentration'.
\end{description}

\vfill
\begin{enumerate}
\item[0/] \textit{lysogeny's means take precedence over a lytic attack $\heartsuit$}
\item[1/] CII determines $\lambda$'s pathway at fresh infection $\heartsuit$: lysogeny at high vs lytic below
\item[2/] the lysogenic cycle is homeostatic, i.e., is sustainable and absorbs perturbations
\begin{enumerate}
\item[a/] protein concentrations are CI controlled in the lysogenic cycle
\begin{enumerate}
\item[$\alpha$/] an initial CI concentration is and must be established by high CII $\heartsuit$
\item[$\beta$/] once established, CI remains physiological, Cro non-physiological $\heartsuit$
\end{enumerate}
\item[b/] a switch to the lytic attack is not effected by
\begin{enumerate}
\item[i/] natural or operative means $\heartsuit$ --- the basis of sustainability
\item[ii/] Cro-perturbation  $\heartsuit$ --- the key means of homeostasis
\item[iii/] super-infection  $\heartsuit$ --- $\lambda$-immunity
\end{enumerate}
\item[c/] a switch to the lytic attack is effected by UV-irradiation (via $\neg\heartsuit$)
\end{enumerate}
\item[3/] the lytic attack expires the host and, else, is sustainable but \textit{not homeostatic}
\begin{enumerate}
\item[a/] protein concentrations are programmatic during a lytic attack
\begin{enumerate}
\item[$\alpha/$] the lytic attack is constitutive $\neg\heartsuit$ or $\spadesuit$
\item[$\beta$/] once initiated, Cro remains physiological and CI inoperative $\neg\heartsuit$ or $\spadesuit$
\item[$\gamma$/] a lytic attack expires the host after Cro has reached high $\neg\heartsuit$ or $\spadesuit$
\end{enumerate}
\item[b/] \textit{ignoring host expiration, a switch to the lysogenic cycle is not effected by}
\begin{enumerate}
\item[i/] \textit{natural or operative means $\neg\heartsuit$ or $\spadesuit$} --- the basis of sustainability
\item[ii/] CI-perturbation $\neg\heartsuit$, although the lytic attack may be set back
\item[iii/] \textit{super-infection $\neg\heartsuit$ or $\spadesuit$} --- an aspect of anti-immunity, see Retrodiction~(V)
\end{enumerate}
\item[c/] \textit{ignoring host expiration, a switch to lysogeny is likely to be effected by}
\begin{enumerate}
\item[ii/] \textit{CI-perturbation $\heartsuit$ (and $\spadesuit$ for the lytic attack to be viable, see 3/d/)}
\end{enumerate}
\item[d/] a lytic attack is unlikely $\heartsuit$ and $\neg\spadesuit$
\end{enumerate}
\end{enumerate}
\caption{$\lambda$\textsuperscript{[AGS3]} ab-intra phenotype (computer-verified)\label{fig:abintra}}
\end{figure}
\begin{theorem}[Ab-intra phenotyping\label{thm:abintra}]
Figure~\ref{fig:abintra} is an \emph{ab-intra phenotype} for \AGSnoun{} organism: \emph{the properties reflect the structure of the reasoning space}, see video. In particular,  the reasoning behind Figure~\ref{fig:abintra} 1) proceeds from \AGSnoun{}'s (regular) genotype, 2) is generated automatically by Figure~\ref{fig:diagram} acting under CCP\textsuperscript{all} on user-given start compartments: the $A_0$s, with termination conditions that generate other $A_0$s, e.g., a few steps or until expiration, looping, or divergence, 3) has been computer-verified to comply with Figure~\ref{fig:logic}, 4) contains \AGSnoun{}'s extant reasoning, and 5) covers \AGSnoun{}'s extant phenotype inferences. \AGSNoun{}'s superseded parts are incompatible with Figure~\ref{fig:abintra} and its $\mbox{justifications}$.
\end{theorem}
The result does not render molecular-biology reasoning trivial. Instead, it shifts the burden to the choice of start points and termination conditions. We make the choices based on what is automatable, with special consideration of real-life events, symmetries, and completeness. The term `ab intra' is due to us but refers to a known phenomenon: axiomatic reasoning is a natural Occam's razor with the effect of streamlining the choice and statement of properties as dictated by the interaction of lower-level issues \citep{Gonthier:4CT08,Nipkow:FAC98,PrincipiaMathematica}. For example, an early milestone in computer-verified reasoning showed that the property progression in the standard reference for axiomatic mathematical reasoning: Whitehead and Russell \cite{PrincipiaMathematica}, allows for a similar kind of automation as here \citep{Wang:MilestoneAward1983}. The wider challenge there at this time is to integrate the in-the-small issue of automation with knowledge-management technologies and user-driven proving strategies \citep{AvigadHarrison:CACM14}.\\

The claim in Theorem~\ref{thm:abintra},1) is the subject of Section~\refSectGenotype{}: MIG/RI specification of \AGSnoun{}, see Appendix~\refAppPremises{}. The scenarios in 2),3) describe use of our tool, see Section~\refSectPhenotype{} and video. The claims in 4),5) are documented in Appendices~\refAppPop{}--\refAppObs{}. The last claim is documented in Appendix~\refAppSups{}: Retrodiction~(I) below contradicts the superseded inferences while Retrodiction~(III) contradicts the superseded assumptions.

\section{\AGSShort{} Retrodiction\label{sect:retro}}
We refer to our formal treatment of \AGSnoun{} organism as $\lambda$\textsuperscript{[AGS3]}.
\begin{description}
\item[(I)] $\lambda$\textsuperscript{[AGS3]} does not require Cro action at O\textsubscript{R}3 against CI, i.e., it goes further in the direction that \AGSnoun{} took the $\lambda$-narrative in its latest edition:
\begin{quote}
\emph{``that Cro must bind O\textsubscript{R}3 to trigger the transition to lytic growth, although not excluded, remains uncertain.''}~\AGSquote{/App.\refAppSups{}}{121}{-4}
\end{quote}
 The quoted triggering was inferred indirectly, based on O\textsubscript{R}3 conservation prior to the recent discovery that an adjacent CI-octamer binding mediates long-distance CI-tetramerization across O\textsubscript{\{L,R\}}3 \citep{Dodd:octa2001,Dodd:coop2004,Revet:CurrBio99}. It is contradicted by Figure~\ref{fig:abintra}:0/, which is a consequence of the discovered CI cooperative binding, see Section~\refSectReasoning{}:0/. See also Section~\refSectExamples{}.

\item[(II)] $\lambda$\textsuperscript{[AGS3]}'s two modes of growth have different stability properties: lysogeny is homeostatic proper; the lytic attack is only so in parts, see Retrodiction~(III), and else is not subject to perturbation, see Retrodiction~(IV).

\item[(III)] $\lambda$\textsuperscript{[AGS3]}'s lytic attack reaffirms that Cro action at O\textsubscript{R}3 is CI-contravariant only late in the attack. The late action increases offspring production:
\begin{quote}
\emph{``If repressor were added to a phage beginning its lytic cycle, growth would be inhibited.''}~\AGSquote{/App.\refAppObs{}}{62}{-1}
\end{quote}
The relevant Cro-affinity is strong but involves non-cooperative binding with a linear effect function: the effect on CI is first neutral then negative.
 
\item[(IV)] $\lambda$\textsuperscript{[AGS3]}'s switch is made efficient partly by not involving Cro action at O\textsubscript{R}3: Figure~\ref{fig:abintra}:0/ implies that proteolysis will have left free CI at zero/sub-physiological concentration once CI's elaborately-cooperative promoter control ceases prior to a switch.

\item[(V)] $\lambda$\textsuperscript{[AGS3]}-variants may exhibit anti-immunity without inhibition of \emph{cII}, cf.:
\begin{quote}
\emph{``The anti-immune phenotype is evidently a consequence of partial repression by Cro of P\textsubscript{L} and P\textsubscript{R} and hence diminished expression of cII and cIII.''}~\AGSquote{}{92}{-7}
\end{quote}
A variant is anti-immune if it can prevent lysogeny by a super-infecting wild type in a contrived lysis-free `lytic cycle'. Weakened Cro binding at O\textsubscript{R}\{1,2\} suffice, see Section~\refSectAntiimmunity{}: higher Cro concentration will be maintained, resulting in continuous CI-contravariance, see Retrodiction~(III). 
\end{description}

\section{Simulation}
\begin{figure}\scriptsize
\begin{center}
\includegraphics[width=\linewidth]{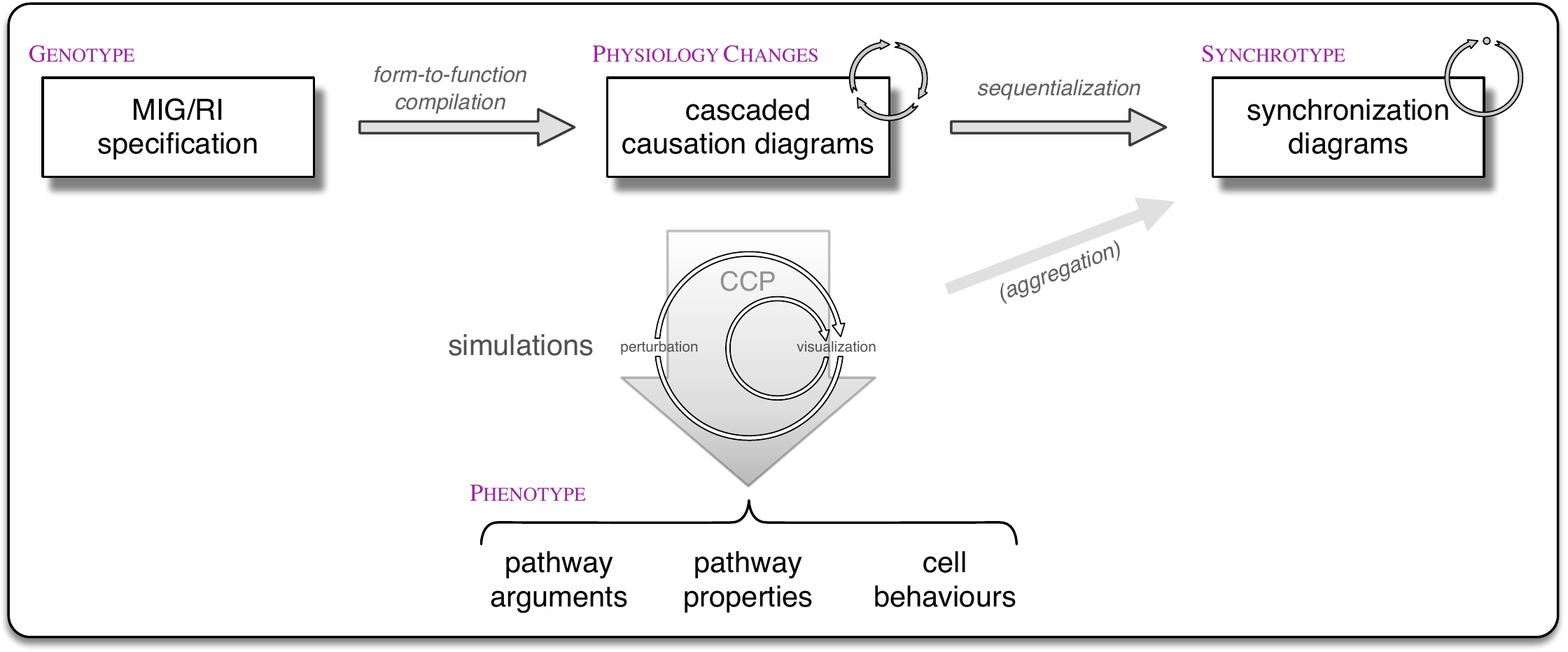}
\end{center}
\caption{Molecular-biology reasoning as computing\label{fig:computing}}
\end{figure}
Summarising the applied side up to this point, we can note that we may ignore the reasoning aspect and view cascaded causation diagrams simply as executable code under CCP\textsuperscript{all}, see Proposition~\ref{prop:automated} and Figure~\ref{fig:computing}.
\begin{proposition}[Open-system concurrent simulation\label{prop:simulation}] Figure~\ref{fig:diagram} \emph{simulates} \AGSnoun{} organism under CCP\textsuperscript{all}: its execution exhibits the considered phenotype properties for the specific reasons from the genotype, including in response to user-imposed differential timing (pursuit of only some enabled causations), environmental signals (external stochasticity: pursuit of edges guarded by non-considered populations), and perturbation (internal stochasticity: pursuit of edges guarded by concentrations other than considered ones). Divergences can be resolved by the user or our tool can do it at random, along multiple strategies.
\end{proposition}
To be clear, we are dealing with three distinct perspectives: 1) real-life mol\-e\-cules and the biology they sustain, 2) the textual molecular biology in \AGSnoun{} that reasons about the connection, and 3) our framework that sits in between the other two perspectives. The text reflects the defining features of the considered real life. Our formalism reifies the text in its entirety. The final question is whether all parts of our formal model is related to the considered real life. The fact that Section~\ref{sect:retro} seems to resolve uncertainties rather than add surprises suggests an affirmative answer. Methodologically, there are three possible sources for any differences: A) the formalism involves principles that do not match real life, B) the formally-specified organism is a poor match for the idealised organism, and C) the idealised organism is a poor match for the real-life organisms, either due to heterogenous populations or poor understanding. Figure~\ref{fig:specification} and Section~\refSectGenotype{} represent our best efforts at addressing B) while C) is partly outside scope and partly tied to A): can our execution complement molecular-biology groundwork, i.e., to what extent is CEq a first-principle framework for real life? Partly to address A) beyond Section~\ref{sect:retro}, we now present derived technologies that involve consideration first of all computation and then of segmented computation that need not reflect the reasoning structure.

\subsection{Sequentialization}
In usage, CCP\textsuperscript{all} execution is best thought of as the interaction of edges as standalone entities rather than the action of a graph on states. `Interacting edges' is a \emph{process calculus} perspective, a class of models of open-system concurrent computation \citep{Milner:TuringLecture93}. Process calculi address not just interleaving or parallel execution of sequential computation but concurrency as a principal notion. Execution in process calculi, including in CCP\textsuperscript{all}, need not have designated beginning or end: we are concerned mostly with them just running, meaning they address computation but not computability per se. Their result notion is external \citep{Mazurkiewicz:Traces77}, as we discuss. Our tool supports the edges-as-processes view by showing implied edges when a node is not used directly but its constituent parts are used separately. The process view provides insights into the workings of an organism and allows us to explore why pathways do or do not exist, see Section~\refSectExamples{}.
\begin{figure}\scriptsize
\begin{center}
\includegraphics[width=\linewidth]{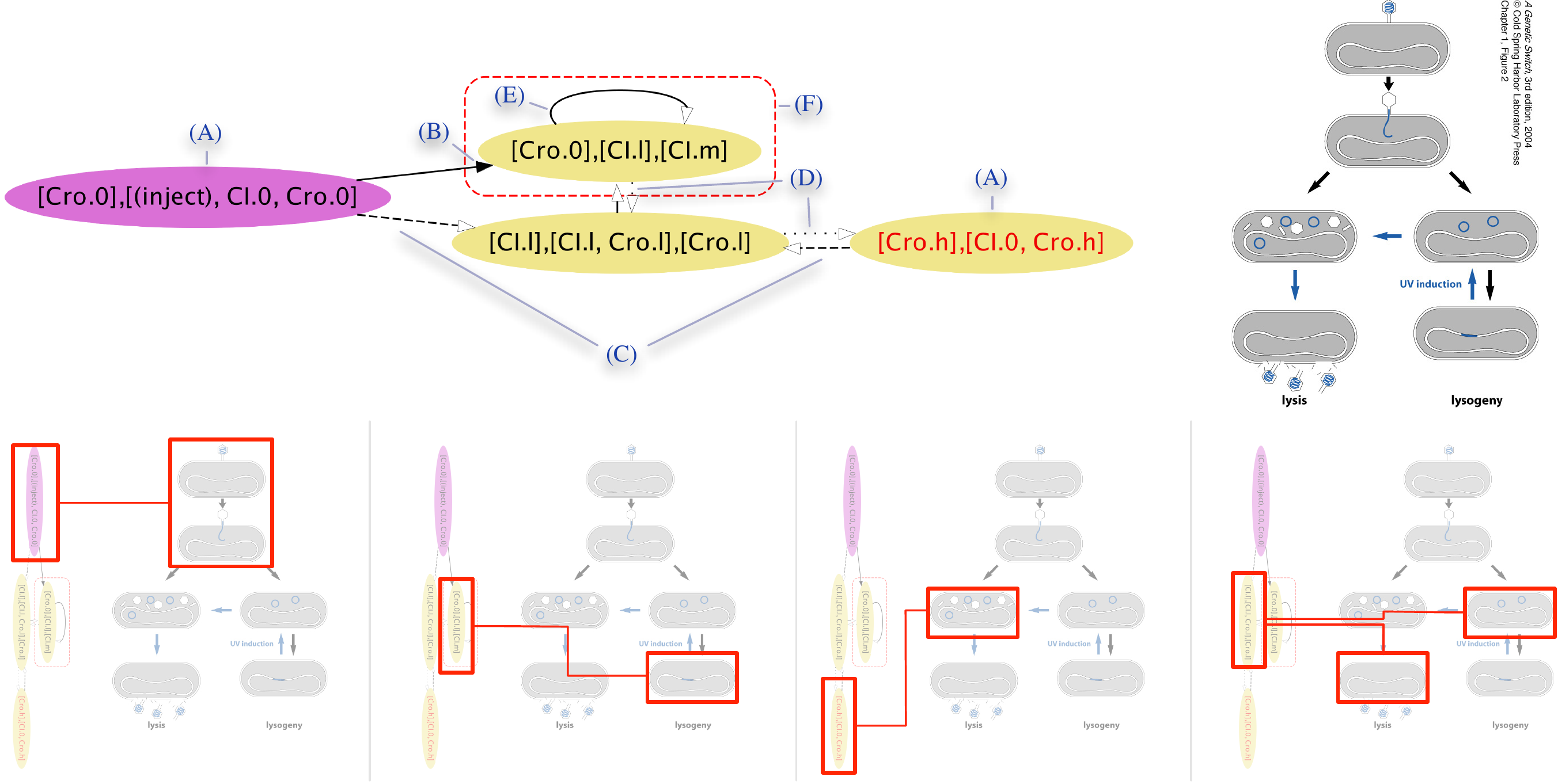}
\end{center}
\caption{$\lambda$\textsuperscript{[AGS3]} \emph{synchrotype} prediction\label{fig:lifecycle}}
\end{figure}
\begin{proposition}[Open-system \emph{synchrotyping}\label{prop:lifecycle}] Our prediction of the possible sequentializations of Figure~\ref{fig:diagram} seen as concurrent CCP\textsuperscript{all} code: \emph{the $\lambda$\textsuperscript{[AGS3]}-syn\-chro\-type}, reifies \AGSnoun{}'s life-cycle illustration, with molecular pathway deciders added, including outside factors, see Figure~\ref{fig:lifecycle}. Synchrotypes operate over \emph{observable} states, found by factoring out `stochastic fluctuations' \cite{Guet:combinatorial2002}.
\end{proposition}
We introduce the name `synchrotype' for the predicted kind of information, to complement `genotype' and `phenotype'. The core of our prediction technology is an adaptation of \emph{trace monoids} as used for process calculi. Conceptually, trace monoids bundle together synchronized processes, leaving only synchrony-breaking changes, i.e., any sequential progress. Technically, trace monoids are to be defined over all linearized coextension derivations (`CCP\textsuperscript{all} traces'), see Figure~\ref{fig:derivations}, by factoring out how causation occurrences may be reordered across interferences, see Section~\refSectSequential{}. The main difficulty here is that causations do not come with explicit handshaking constructs that control synchronization/reordering. For that, we need to integrate inhibition into cascading, etc. For example, the lytic-attack processes (\emph{cro}-transcription) can only come after those that maintain lysogeny (\emph{cI}-transcription from P\textsubscript{RM}) because CI at O\textsubscript{R}\{1,2\} effects synchronization by simultaneous inhibition respectively auto-mediation. Our prediction technology applies to cascaded causation diagrams, with matching motivation. In principle, we collapse looping edges and take intermediates of the affinity-transposing states in the connected nodes as `observable states':
\begin{quote}
\emph{``As [Cro] binds it turns off [P\textsubscript{R}], but as the cells grow and divide, the concentration of [Cro] drops, and [P\textsubscript{R}] turns on again. A steady state is reached at which the rate of synthesis of [Cro] just balances its rate of dilution and, presumably, a constant concentration of [Cro] is maintained. In this situation [Cro] diminishes (turns down), but does not abolish, its own synthesis.''}~\AGSquote{/App.\refAppAxioms{}}{92}{-3}
\end{quote}
The loops/`observable states' are tested for coexistability under inhibition in order to scale up to compartment-wide observations and the considered cascaded causation diagram is correspondingly convolved as the final step. The difficulty is that `loop' is not a simple notion for edges with negative conditions under coexistability. Instead, we introduce \emph{sustained equilibria} in place of plain loops. In graph theory, a \emph{strongly-connected component} is a collection of nodes within a graph such that 1) there is a directed path of edges from any node to any other node and 2) no node can be added to the collection without breaking 1).
\begin{definition}\label{defn:sustained}
Given a classification of states as A(lways)/T(ran\-si\-to\-ry)/N(ever) sustainers, we find all strongly-connected components over the graph without A-inhibited edges. We then classify any found component as a \emph{sustained equilibrium} of the given ATN type if it has no out-edges in the graph without A- and/or T-inhibited edges. By default, we consider only direct inhibitors for sustaining: indirect inhibition tends to express precedence, i.e., other action dominates.
\end{definition}
The default definition of coexistability, below, is too strict. It, e.g., does not allow for lock-stepped interleaving, but it mostly suffices for the monograph. \begin{definition}\label{defn:coexist}
A collection of sustained equilibria are \emph{orthogonal} if no schema instance is in different states in different equilibria and no equilibrium loses strong-connectivity under inhibition by the content of the other equilibria.
\end{definition}
We shall not pursue the issue of a perfect definition here, other than mention that our tool also comes with a notion that is too lax, called \emph{reconcilable} equilibria: each node in an equilibrium must be able to pair up with a node from each of the other equilibria without having some schema instance be in two states.\\ 

The trace-monoid perspective on our synchrotype prediction is that processes are nominated as independent of each other, i.e., reorderable, first within sustained equilibria, Definition~\ref{defn:sustained}, and then across coexistable equilibria, Definition~\ref{defn:coexist}. If an ATN-sustained equilibrium cannot coexist with another, it is possible a subsumed one can, with sustainers reclassified from N to T to A.

\subsection{Rapid Phenotyping} 
In addition to reasoning about the wild type's \emph{temperate} phenotype, \AGSnoun{} includes counterfactual reasoning about mutants exhibiting \emph{clear}, \emph{virulent}, and \emph{anti-immune} phenotypes, with the latter needing a host that cannot lyse. The \emph{clear} phenotype differs from \emph{temperate} by pursuing a lytic attack on all fresh infection, see Figure~\ref{fig:abintra}:1/. The \emph{virulent} phenotype additionally overrides the wild-type's ability to withstand super-infection, see Figure~\ref{fig:abintra}:2/b/iii/. The \emph{anti-immune} phenotype differs from \emph{temperate} by being able to prevent lysogeny from a super-infecting wild type, see Figure~\ref{fig:abintra}:3/c/ii/. Inspired by the automation in the wild-type's ab-intra phenotype, see Figure~\ref{fig:abintra}, a few concrete tests can be used to rapidly phenotype variant organisms, see Section~\refSectRapid{}: what is the outcome from fresh infection, can Cro's concentration increase in the presence of CI, what is the outcome from having CI at its highest and Cro at zero concentration, what is the outcome from there of raising the observed Cro concentration respectively introducing short, single-stranded DNA, and what is the outcome from having Cro at its highest and CI at zero concentration (with expiration turned off in the tool options). Section~\refSectAbstracting{} presents a generic MIG-specification that allows us to vary the intrinsic affinities of the proteins for the operators, for 1,728 variants. The interactive analysis of each variant takes less than a minute. The initial challenge is to account for alternate-pairwise CI cooperativity, a form of meta-regulation:
\begin{quote}
\emph{``Because repressor dimers at O\textsubscript{R}1 and O\textsubscript{R}2 interact, or repressor dimers at O\textsubscript{R}2 and O\textsubscript{R}3 interact, we say the cooperativity is `alternate pairwise'.''}~\AGSquote{}{21}{-2}
\end{quote}
1) Our MIG language is sufficiently abstract that it can accommodate within a single specification all the regulation changes that result from varying the intrinsic affinities, i.e., MIG covers also the considered meta-regulation, see Section~\refSectAbstracting{}. 2) Our RI modeling is sufficiently robust that the generic specification needs only local changes from the wild type, see Figure~\ref{fig:specification}. 3) A caveat: \AGSnoun{} does not discuss all changes to the highly-cooperative molecule interactions that may result from varying the intrinsic affinities and our analysis is based in part on surmises. An example concerns favored CI binding to O\textsubscript{R}3 over O\textsubscript{R}1, where the wild-type's favored cooperativity direction becomes moot. We surmise that the resulting long-distance octamer does not allow for a subsequent long-distance tetramer on the free O\textsubscript{\{L,R\}} operators, i.e., the discovery that prompted the latest \AGSshort{} edition \citep{Ptashne:AGS3}. We imagine that the operators will not be sitting across from each other unless a strand of the super-coiled DNA is in the opposite orientation of the wild-type case, see Retrodiction~(I). 4) We have seemingly identified a new \emph{anti-immune} variant, see Retrodiction~(V) and Section~\refSectAntiimmunity{}.
\begin{figure}\scriptsize
\begin{center}
\includegraphics[width=\linewidth]{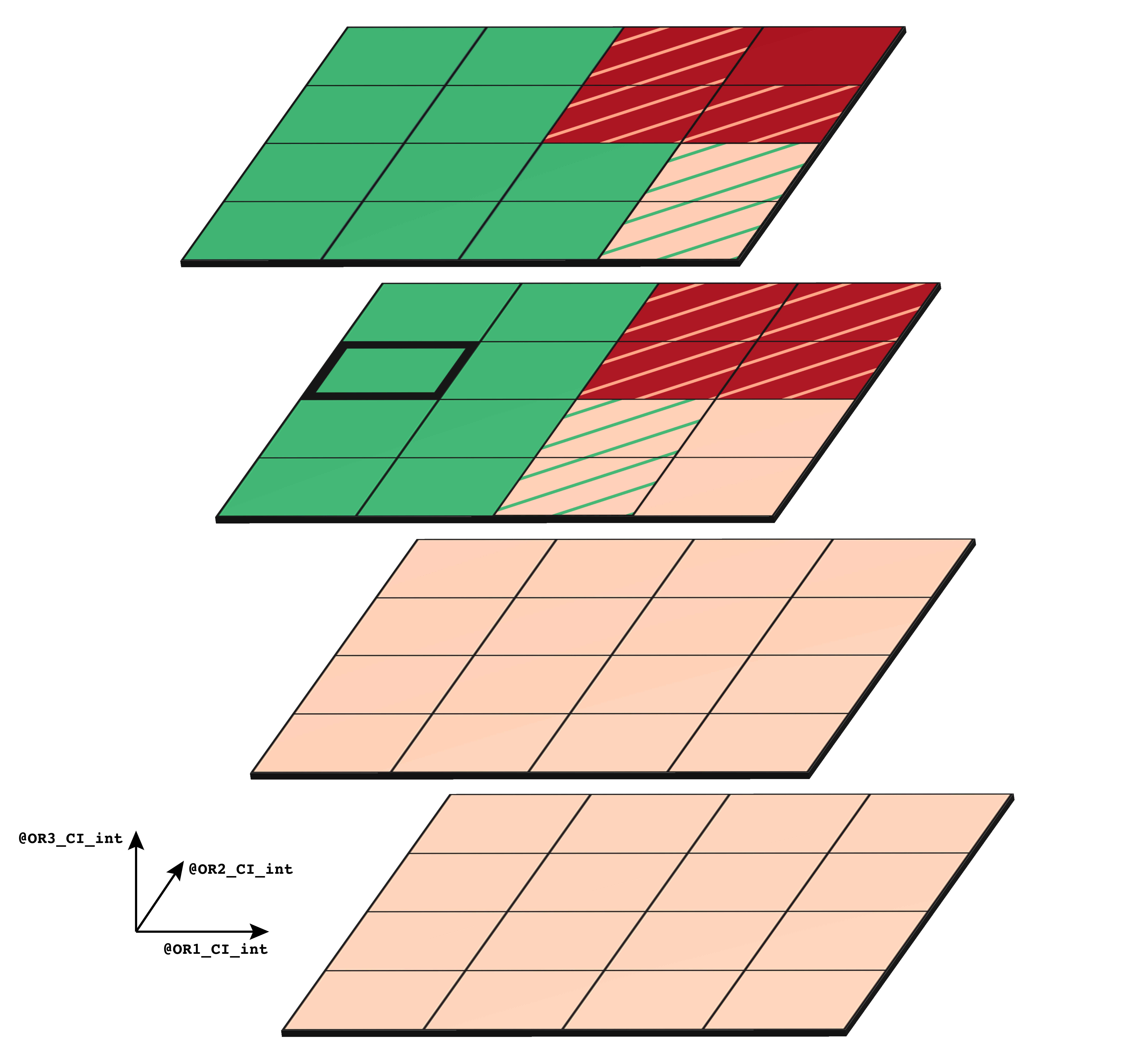}
\end{center}
\caption{Phenotype phase space for $\lambda$\textsuperscript{[AGS3]}-mutants: varying CI affinities\label{fig:phasespace}}
\end{figure}
5) Figure~\ref{fig:phasespace} shows the phenotype phase space from varying CI's intrinsic operator affinities, with \emph{temperate}, \emph{clear}, \emph{virulent}, and hybrids, see Section~\refSectCIVariants{}. The \emph{temperate}-\emph{clear} hybrid codes for a divergence: alternate-pairwise cooperativity may go either way, suggesting that the phenotype is chosen stochastically. However, the choice gets made repeatedly and the variant will probably manifest as \emph{clear}. Higher than wild-type CI affinity for O\textsubscript{R}3 (bottom of Figure~\ref{fig:phasespace}) results in the \emph{clear} phenotype: CI works against CI-maintained lysogeny. Otherwise, the exhibited phenotype will tend from \emph{temperate} through \emph{clear} to \emph{virulent} as it becomes increasingly difficult for CI to inhibit \emph{cro}-transcription (front-left to rear-right in Figure~\ref{fig:phasespace}). To the extent of the available mutation information, Figures~\ref{fig:abintra},\ref{fig:phasespace} are similarly compatible with \AGSnoun{}, see Section~\refSectCountercorrect{}. An alternative explanation for the prevalence of \emph{clear} over \emph{virulent} variants (other than involving fewer mutations \AGSquote{}{68}{13}) is that mutations that increase CI's O\textsubscript{R}3 affinity trump other mutations affecting CI's affinities in terms of the exhibited phenotype.

\section*{Discussion}

``Among the sciences, mathematics is distinguished by its precise language and clear rules of argumentation'' \citep{AvigadHarrison:CACM14}. We establish that molecular biology may be undertaken in the same manner, with several derived benefits. Fundamentally, we have verified that the standard molecular-biology reference \citep{Ptashne:AGS3} is \emph{formally correct}: its reasoning complies with a consistent logic, i.e., it makes sense, technically. Rather than relying on existing technology for showing formal correctness, we have recapitulated the modern pillars of mathematical reasoning:
\begin{description}
\item[\emph{Principia Mathematica} \citep{PrincipiaMathematica}]$\ $ axiomatized mathematical practice and gave rise to the standard for rigor of the 20th century. \emph{Principia Mathematica} is \#23 in ``The Modern Library's Top 100 Nonfiction Books of the Century,'' see {\footnotesize\texttt{https://www.nytimes.com/library/books/042999best-nonfiction-list.html}}.

\item[The Curry-Howard Correspondence \citep{Howard:CurryHoward,Wadler:PropTypes2015}] says that mathematical practice is closed-system sequential computation and vice versa. We establish that mo\-le\-cu\-lar-biology practice is open-system concurrent computation. Math\-e\-ma\-tics-type computation can mimic the molecular-biology type up to termination, the presence of unresolved choices, and stochasticity.

\item[``The new standard for rigor'' \citep{AvigadHarrison:CACM14}] looks set to become computer-assisted reasoning with verification and development technology, using axiomatics and proof by reflection (e.g., the innate Curry-Howard form).

\item[Constructivity \citep{Bauer:AcceptingConst2017}] guarantees that molecular-biology reasoning is molecule-based simulation of the biology. Generally, constructivity is relative to a language of properties, meaning we have alignment of concerns when there is language matching at all levels. Reductionist reasoning is (all-level) constructive by definition. See below for the non-constructive case.
\end{description}
The outcome here is ``a theory of constructive engineering principles of life [over molecular interactions]'' \citep{US-NRC:TheoryBiology}, i.e., first principles from reductionism. This should be understood in contrast to Anderson \citep{Anderson:MoreDifferent1972}: while reductionism per se is not constructionism, reductionist \emph{reasoning} is. The induced theory consists of:
\begin{description}
\item[Molecular programming:] MIG/RI-specifications capture how molecules may interact: RI stands for Regulation Interface; states nominally transpose affinities, i.e., they capture when bindings may have material effects; regulation may be arbitrarily nested; regulatory units may be assigned a name for subsequent tracking in the reasoning; and modalities may be used to account for variations in the manifestation of the programming, e.g., whether a mediator needs to remain in place to see a change through.

\item[Causations] are constructed as the independently-operating and discretely-reg\-u\-la\-ted manifestations of the molecular programming, accomplished as conversion of propositional versions of regulatory expressions to disjunctive normal form. Causations are `elementary processes of physiology change'.

\item[Concurrency:] Everything biological is inferred by assessing how causations may interfere as concurrent processes. The interference works as a process calculus, a type of computation that need not have designated beginning or end and where no particular form of input or output is involved: the computation just runs as it reacts to any and all environment changes. All runs are guaranteed to be coded for by justifying molecular interactions.

\item[Sequential results:] An external result notion exists that aggregates how the processes can interfere, by factoring out how they may be reordered across interferences. The notion captures the possible sequential forms the concurrency may take: the coded-for life cycles/the organism `synchrotype'.

\item[Open-system modeling] involves weaker logical assumptions than a closed-sys\-tem approach, meaning an open-system framework will be sound for more real-world scenarios. Here, closedness admits false-positive argumentation for properties that hold for different reasons, see below. Aside from avoiding state spaces and appeals to truth within our setup, open-system modeling dictates that we do not enforce inhibition when predicting synchrotypes but retain the possibility as a pathway decider, see Figure~\ref{fig:lifecycle}.

\end{description}
While the induced mode of computation will not account for all of molecular biology, it does involve abstract concepts that go beyond \AGSnoun{} \citep{lambdaPhantom} and seemingly are needed throughout the discipline \citep{Ptashne:NatChemBio11}: we establish the basics.
\begin{description}
\item[Hybrid modeling:] Owing to coding: the absence of computable-violating argumentation in CEq, it is possible to instrument a genotype with affinity values and push them through to kinetics values on the edges in cascaded causation diagrams alongside the regulatory-expression names that track the sites of action, see Constructivity Revisited below. In what seemingly is not a surprise \citep{Einstein:GeoExp1921}, numerical and symbolic methods that rely on and respect the same reductionist backbone can be expected to combine:
\begin{description}
\item[ab initio:] first-principle numerical calculation over physical properties of base entities.
\item[\emph{ab intra:}] \emph{first-principle symbolic computation over logical relationships between base entities.}
\end{description}
Combining would use ab-intra technology for the backbone and ab-initio ones for specialized tasks, e.g., a more realistic CCP-strategy than `all'.

\item[Validating prediction:] The key concept in our reasoning and computation formalisms is causations. As detailed above, causations are constructed as exactly the independently-operating processes of physiology change, and we predict that they can be identified in biology, too. Based on independence, we may detect them combinatorially. One candidate scenario is digitation. The hypothesis would be that each causation resulting from the regulation of a growth factor is responsible for the growth of one digit. Figure~\ref{fig:causations}:(E) shows that the open issue of Fibonacci-many digits might be explained by nested inhibition but the prediction is more general. 

\item[Complexity:] Our translation of the regulation of a gene into the coded-for causations may need to consider as many as double-exponential-many potential causations in the size of the start expression. If we abandon the requirement that causations operate independently of each other, i.e., if we admit false-positive argu\-mentation/site-action, some exponential-many actual causations collapse to linear-many, see Figure~\ref{fig:causations}:(E). The complexity for a genotype is the summation over the individual genes. Further to the preceding item, we note that the complexity has direct bearing on the C-value enigma (no correlation between real-world genome size and organism abilities): the dominant asymptotic driver is regulation.

\end{description}

Our approach is \emph{intensional}: we directly address regulatory engineering principles over mature concepts, with effects retained as effects. Existing approaches to the analysis of gene regulation use state spaces and are \emph{extensional}: they address effects throughout. As shown, life cycles, etc., are intensional notions. In general, intensional notions (e.g., physiological concentrations) are not easily recovered from extensional ones (e.g., observable concentrations), see Section~\refSectDiagrams{}.\\

Given proven reasoning principles, logical meta-theory makes fairly easy work of engineering structure and enables computational treatment of dynamics. Our axiomatization of molecular biology has been tested in exacting detail:
\begin{itemize}
\item It involves abstract principles \citep{lambdaPhantom} that are subsumed by standard reasoning, see Theorem~\ref{thm:coding}.
\item It has been computer-verified to enjoy standard reasoning meta-theory, see Theorems~\ref{thm:consistency},\ref{thm:consequence}, Corollary~\ref{cor:cut}.
\item It has been computer-verified to account for the extant parts of the standard reference \citep{Ptashne:AGS3}, see Theorem~\ref{thm:abintra}.
\item It admits derived usages covering all rather than just specific reasoning, including automation, see Theorem~\ref{thm:abintra}, Proposition~\ref{prop:simulation}.
\item Its derived usages admit their own derived usages, see Proposition~\ref{prop:lifecycle}, \citep{US-NRC:TheoryBiology}.\\
\end{itemize}
The induced theory appears to be simple. This is a reflection of conceptual purity and is a hard-won property. Without simplicity, there would be caveats in our story for Ptashne \cite{Ptashne:AGS3}. As seen, the theory's inner structure is non-trivial.

\subsection*{Constructivity Revisited}

As presented, `constructivity' is a requirement of an axiomatic proof system over the given formal language: \constructive{}. The most interesting instances are called the disjunction and existence properties after the connectives in question. The law of excluded middle (LEM: $p\vee\neg p$, for any $p$) violates the disjunction property, i.e., constructivity: LEM does not mandate a proof of either $p$ or $\neg p$.
\begin{quote}
\emph{``Taking the principle \emph{[here: `law']} of excluded middle from the mathematician would be the same, say, as proscribing the telescope to the astronomer or to the boxer the use of his fists. To prohibit existence statements and the principle of excluded middle is tantamount to relinquishing the science of mathematics altogether.''}~\citep{Hilbert:GrundlagenMath1928}
\end{quote}
The logic commonly used for mathematical reasoning is called classical logic. It is obtained by adding LEM to intuitionistic logic, see Theorem~\ref{thm:coding} and Section~\refSectFormalReasoning{}. Adding LEM makes more properties provable and, in effect, collapses provability to an external notion of truth: a proposition is classically provable if and only if all entries in its truth table evaluate to true. With this, LEM allows the classical reasoner to think in terms of properties rather than justifications. For a direct characterization, we can note that LEM is equivalent to double-negation elimination ($\neg\neg p\rightarrow p$, for any $p$) over intuitionistic logic, meaning a property, $p$, got added on the grounds that it might as well be true in the sense of not violating consistency: $\neg\neg p$, or because it is dependent on such $p$s.

No notion of truth is involved in constructive systems, implicitly or explicitly, corresponding to the scenario that we may not yet know if either a formula, its negation, or neither can be justified from the ground up. Constructively, formulas do not have and cannot be given a truth or a truth-like status. Glivenko's Theorem tells us that a propositional formula is provable in classical logic if and only if the double-negation of the formula is provable in intuitionistic logic (with similar results for other logics and/or formula encodings), meaning constructive reasoning is more discerning but not less expressive than truth-based reasoning. The view espoused by Hilbert \cite{Hilbert:GrundlagenMath1928} is now held less strongly.\\

Within classical reasoning, LEM can be restricted to variables (aka stability) without loss of conclusions, see Section~\refSectFormalReasoning{}. For scientific reasoning, where all variables refer to external notions, e.g., concentrations, stability amounts to a closed-system assumption: the external entities are always either present or absent. This means that reasoning over state spaces is equivalent to classical reasoning: no matter how discerning you attempt to be, constructivity's inherent guarantee for ground-up justifications: `coding', has been undermined. The severity of this depends on how the non-constructive principles get employed. State spaces could have been used in CEqEA 1) for inferring causations from genotypes, see Sections~\refSectDNF{},~\refSectPhysiology{}, and 2) for visualizing causations, see Sections~\refSectElementary{},~\refSectCoextSS{}. For 1) alone, we would violate not just Theorem~\ref{thm:coding} as stated but also its intention: as it stands, macro names created by MIG/RI specification, see Figure~\ref{fig:specification}, show up in the right places in subsequent reasoning, i.e., we reify the monograph reasoning, see Figure~\ref{fig:derivations}. This would no longer be possible without pervasive false positives. Technically, false-positive names would undermine coinhibition and mutually-exclusive binding, see Section~\refSectAbstracting{}. For 2) alone, the practical automation features we have added in CEqEA, see Sections~\refSectPerturb{},~\refSectInter{}, would be difficult to accommodate and false-positive divergences get introduced for Section~\refSectReasoning{}:2/b/ii/,3/b/ii/, see also Section~\refSctSynchroDgrm{}. Practically, classical reasoning/state-space modelling would use both 1) and 2).

\subsection*{Related Technologies (MINIMALIST VERSION!)}

The only related work proper is interactive theorem proving for mathematics with the Curry-Howard Correspondence and attendant types of applications \citep{AvigadHarrison:CACM14}, as discussed. Owing to our vertically-integrated work, we touch on many specific subjects. Each of the involved methodologies and technologies is important, but our overarching contribution is their integration: reductionist reasoning as first-principle simulation. All specific contributions are dictated by the way in which the involved methodologies and technologies are being integrated.\\

For an example of a specific technology contribution, see this: 
\begin{quote}
\emph{``Today, mobile phones, server farms, and many-core processors make us concurrent programmers. [\ldots] \textbf{Many process calculi have emerged} --- ranging from Communicating Sequential Processes (CSP) to Calculus of Communicating Systems (CCS) to $\pi$-calculus to join calculus to mobile ambients to bigraphs --- \textbf{but} [\ldots] \textbf{none has the distinction of arising from Curry-Howard}.''}~\citep{Wadler:Session-JFP2014}
\end{quote}
Our Calculus of Coextensive Processes (CCP) is a such-arisen process calculus.

\newpage\section*{Figure Legends}

\begin{description}
\item[Figure~\ref{fig:specification}:] A cross-section of the formal $\lambda$\textsuperscript{[AGS3]}-genotype, covering regulation of \emph{cro}-expression. The text is in our MIG (Modal Influence Graph) language, written in accordance with our RI (Regulation Interface) modeling methodology applied to Ptashne \cite{Ptashne:AGS3}, see Section~\refSectGenotype{}. The combination is called MIG/RI specification. The items are listed in reverse for discussion purposes. Macros, for regulatory expressions, are prefixed with \texttt{@} and categories with \texttt{\$}. Conjunction is written \texttt{\&}, disjunction \texttt{:}, inhibition \texttt{|-{}-}, and contravariance \texttt{!}. The text inside {\texttt{!\ \{|\ldots\}}} is a modality for the contravariant effect. A \texttt{|} is a separator and a \texttt{;} is a terminator. Comments are after \texttt{//} --- here, they are used to indicate the main quotations in Ptashne \cite{Ptashne:AGS3} being formalized under RI.
(A) The \emph{cro} gene is transcribed from the P\textsubscript{R} promoter, which increases (\texttt{+}) the concentration of Cro protein; Cro decays naturally and is subject to passive contravariance (\texttt{!}): the population decreases unless maintained. 
(B) P\textsubscript{R}-transciption is constitutive but inhibited by either of the O\textsubscript{R}\{1,2\} operators; Cro has a neutral effect at one site at first but is contravariant at higher concentrations (\texttt{+-}) and at dual occupancy (\texttt{--}); CI binding is cooperative, immediately contravariant, and may outlast its free concentration ({\texttt{\{|\ldots\}}}). 
(C) Dual Cro occupancy of O\textsubscript{R}\{1,2\} happens at the highest concentration, if both are well-defined (\texttt{\&}); single occupancy requires either to be well-defined (\texttt{:}) and is \texttt{excl}uded by dual. 
(D) Cro regulates with concentrations matching \texttt{int}rinsic affinities but CI binds preferentially. 
(E) Highly-cooperative CI binding is either as an octamer at  O\textsubscript{\{L,R\}}\{1,2\} plus a tetramer at O\textsubscript{\{L,R\}}3 or as just an octamer. 
(F) CI's octamer binding at O\textsubscript{\{L,R\}}\{1,2\} (\texttt{@OLR12\_CI}) is initiated from O\textsubscript{R}1 and the cooperativity makes it physiological at a lower concentration (\texttt{(+) ...}) than that transposing the \texttt{int}rinsic affinity; octamer+tetramer binding forms when CI is able to bind also at O\textsubscript{R}3, with cooperativity across O\textsubscript{\{L,R\}}3 lowering the effective concentration. 
(G) CI binds O\textsubscript{R}\{1,3\} with different \texttt{int}rinsic affinities, specifically at low and high nominal concentrations.
(H) Cro binds O\textsubscript{R}\{1,2\} at high nominal concentration.
(I) \emph{cI}/CI and \emph{cro}/Cro are declared as genes whose protein products bind operators differentially, at several nominal concentrations.

\item[Figure~\ref{fig:logic}:] Axiomatic compartment changes over causations, i.e., over the dis\-crete\-ly-regulated physiology changes coded for by a genotype, see Figure~\ref{fig:causations} and Section~\refSectPhysiology{}. In Definition~\ref{defn:pos_cstn}, $\doteq$ means `defined to be', see Section~\refSectIndDefn{}. Definition~\ref{defn:coextension} is an inductive definition of a formal proof system by proof rules, see Section~\refSectIndDefn{}: `if above the line, then below', with $g\vdash^{\mbox{\scriptsize co}} A_a{\hookrightarrow}A_b$ meaning: `we [can] infer $A_a{\hookrightarrow}A_b$ from $g$'. The $\{R_i \stackrel[I_i]{M_i}{\rightarrow} P_i\}$ are user-chosen, indexed subsets of the po\-si\-ti\-ve\-ly-va\-li\-da\-ted causations in the considered compartment. The $R_i \stackrel[I_i]{M_i}{\rightarrow} P_i$ must be (co-)inhibition-free and produce a homogeneous compartment when combined. We consider open systems: mediators and inhibitors need not occur as reactants or products.

\item[Figure~\ref{fig:causations}:] (A) The causations coded for by Figure~\ref{fig:specification} (but without macro names, see Figure~\ref{fig:derivations}) --- for the specifics, see next. See Section~\refSectPhysiology{} for all causations.
(B) The `causation arguments' in Ptashne \cite{Ptashne:AGS3} typically synthesize information spread over multiple causations. Reading (A) in textual order, \AGSquote{/App.\refAppCArg{}}{22}{1} makes reference to the inhibition on the first two causations: constitutive \emph{cro}-expression. With the first two, the third causation completes \AGSquote{/App.\refAppCArg{}}{25}{7} while stressing that the referenced operator bindings are contravariant, see Figure~\ref{fig:specification}:(B): they result in an opposite effect, meaning effective decay. The last six causations are the result of contravariance from \AGSquote{/App.\refAppCArg{}}{22}{1}. More subtly, \AGSquote{/App.\refAppCArg{}}{22}{1}'s ``[i]n a lysogen'' is seen as the CI inhibition on Cro-mediated auto-decay in the third causation: CI and Cro bind to the same operators to mediate Cro-decay, but CI binds preferentially, see Figure~\ref{fig:specification}:(D). Without the CI inhibition, our framework would admit false-positive argumentation for effects that do take place but for different reasons.
(C) The essence of our translation from the MIG language, see Figure~\ref{fig:specification}, to causations, see (A), is to view regulatory expressions as propositional formulas ($\lceil\_\rceil$) and convert these to \emph{disjunctive normal form} (2DNF) by using commutative laws to push all negations to the inside, conjunctions to the middle, and disjunctions to outermost, see Section~\refSectFormFct{}. Each non-contradictory disjunct (i.e., maximal collection of conjunctions of arbitrarily-negated states) is then the conditions for a causation. A key aspect is the listed definition of propositional inhibition. The obvious definition may appear to be $\lceil\texttt{e1}\rceil\wedge\neg\lceil\texttt{e2}\rceil$, with the listed example instead becoming $(o_2.s\wedge \neg o_1.s)\vee(o_2.s\wedge\neg\neg o_0.s)$. The first of these two disjuncts would become a causation that can go ahead in the presence of $o_0.s$ although $o_0.s$ is stated to inhibit the considered inhibition by $o_1.s$. The second can also go ahead, but the first would be false-positive argumentation. The definition we give ensures that all considered causations operate independently of each other.
(D) Consider schemas, $o_i$, with one active ($s$) and one non-active state each. Let $\mathrm{NI}(n)$ be the function that constructs $n$-nested inhibition over the $o_i.s$ by \emph{structural recursion}, see \refSectIndDefn{}.  The example in (C) is $\mathrm{NI}(2)$. 
(E) Partly to illustrate the effective structure resulting from (C), we prove that $\mathrm{NI}(n)$ gives rise to Fibonacci number ${n+1}$ ($\emph{F}_{n+1}$) many causations. The base cases involve $o_0.s$ and $o_1.s\wedge\neg o_0.s$. The proof steps for the recurrence are (1) definition of inhibition; (2) standard distributive laws for conjunction over disjunction, Section~\refSectDNF{}; (3) the conjunctive clauses in a DNF is the union of clauses over a disjunction; (4) conjoining a variable onto an formula does not change the number of clauses and, as it is a distinct variable, no causations get invalidated; no causations result from a contradiction; (5) by definitions; propagate negation by De Morgan's laws (DM), Section~\refSectDNF{}; (6) repeat step in (3); (7) repeat step in (4); (8) twice DM leave disjunctions and conjunctions intact. Use (4)$\mapsto$[(1) left-hand] twice in [(5) left-hand]=(8). For propositional inhibition as $\lceil\texttt{e1}\rceil\wedge\neg\lceil\texttt{e2}\rceil$, the result for $\mathrm{NI}(n)$ becomes ${\mathrm{IntegerPart}(n/2)+1}$. $\emph{F}_{n}$ is a bit more than 2$^{n/2}$.

\item[Figure~\ref{fig:derivations}:] Consider 
(A) the listed set of states, with instance-sorting given by the part before `.', and 
(B) these `[named]:' causations coded for over them, 
where (C) [\texttt{SOS}] coinhibits (outpaces) [\texttt{PRM\_tr}] --- the listed causations are Section~\refSectLambdaTrans{}: 1a1:1;2.1:3;2.1:4;3.2:2;4.1:1 for $\lambda$\textsuperscript{[AGS3]}. 
(D) An example coextension derivation, see Figure~\ref{fig:logic}, with rule names omitted: (interference) occurrences are top-most and include positively-validated causations in $[\![\ldots]\!]$, with (co-)inhibitees to the right of semi-colons and the rest used for $\{R_i \stackrel[I_i]{M_i}{\rightarrow} P_i\}$.
(E) Our tool outputs coextension derivations in linearized form, with enabled positively-validated causations after \texttt{->} or, if (co-)inhibited, after \texttt{x>}. \emph{Inert} states not used as reactants are written after \texttt{+}. Our linearized coextension derivations reify the `pathway arguments' in Ptashne \cite{Ptashne:AGS3}. For the example, \AGSquote{/App.\refAppPArg{}}{24}{-7}'s ``SOS response'' is listed in l.2; it is effected by (activated) RecA* in l.1; SOS counteracts \emph{cI} expression from P$_{RM}$, see l.3, which results in a CI-concentration decrease: l.1 vs.\ l.5; ``vacating repressor'' (CI) is \texttt{OLR12\_CI} occurring in l.3 but not in ll.5--7; ``transcription of \emph{cro}'' from P$_R$ is listed in l.6, resulting in a Cro-concentration increase: l.5 vs.\ l.9. See video.

\item[Figure~\ref{fig:diagram}:] The graph of the coded-for causations from the $\lambda$\textsuperscript{[AGS3]} genotype, see Section~\refSectCCD{}. The darker nodes in boxes list schema instances that do not change across any edges, i.e., environmental influencers. The magenta node with (inject) was specified as an entry seed in the $\lambda$\textsuperscript{[AGS3]} genotype: a channel into the considered compartment, see Section~\refSectSpecials{}. The edge labels show regulators and where they act: \texttt{<source-only> ; <source+target> ; <target-only>}. An edge with a filled head has a target-only mediator, which must be present for the edge to run. A \texttt{-}/$\sim$ indicates direct/indirect inhibitors. A \texttt{!} indicates inhibitors with a contravariant effect that will appear elsewhere. The red text in the \texttt{Cro.h} node indicates system expiration (here: host lysis), see Section~\refSectSpecials{}.

\item[Figure~\ref{fig:abintra}:] Our formally-substantiated `pathway properties' for Ptashne \cite{Ptashne:AGS3} are developed \emph{ab intra}, see Theorem~\ref{thm:abintra}: they reflect the structure of the reasoning space generated by Figure~\ref{fig:diagram} acting on compartments, see video. The listed \emph{$\lambda$\textsuperscript{[AGS3]} phenotype} and its reasoning cover Ptashne \cite{Ptashne:AGS3}, see Theorem~\ref{thm:abintra}. The items in \emph{italics} are not in Ptashne \cite{Ptashne:AGS3}, see Retrodictions.

\item[Figure~\ref{fig:computing}:] The practical aspect of this work is a tool that supports 2 usage perspectives: molecular-biology [reasoning $\leftrightarrow$ computing]. The use case of Genotype [molecular basis $\leftrightarrow$ \emph{programming}] to Physiology Changes [`causation arguments' $\leftrightarrow$ \emph{executable code}] to Synchrotype [\emph{open-system} life cycles $\leftrightarrow$ \emph{(aggregated) sequential form}] is automated. Exploration of Phenotypes [`pathway arguments' + `\emph{pathway properties}' + `cell behaviors' $\leftrightarrow$ \emph{perturbable concurrent computation}] is interactive. The affixed circles indicate the computational nature of the locations: physiology changes operate concurrently; the simulations may effect these step-wise; synchrotypes  factor out how the concurrent processes may synchronize to each other in a stepping-free way, resulting in a sequential presentation of all possible behaviors, i.e., a prediction of the possible system life cycles.

\item[Figure~\ref{fig:lifecycle}:] Our prediction of $\lambda$\textsuperscript{[AGS3]}-sequentializations reifies the temperate life-cycle illustration in Ptashne \cite{Ptashne:AGS3}, see \AGSquote{/App.\refAppObs{}}{133}{15}. The construction convolves relevant parts of Figure~\ref{fig:diagram}, with originating nodes in []. The obtained graph reveals molecular details that would affect its traversal.
(A) Synchronization nodes consists of coexistable causation nodes, with channels (magenta) and expiration (red text) from the causation level. The expiration node is not terminal because lysis is not regulatorily terminal, see Retrodiction~(II).
(B) Edges with a filled arrow have a target-only mediator: the direct route to lysogeny is available only to a super-infection in the considered case of CII below high concentration, see Figure~\ref{fig:abintra}:1/,2/b/iii/. It is not filled for CII at high, see Section~\refSectAGSsynchro{}.
(C) Dashed edges are inhibited in the target node. CI is an inhibitor for the left edge, i.e., only continued lysogeny is possible in case of super-infection, see Figure~\ref{fig:abintra}:2/b/iii/. See Retrodiction~(III) for the right edge.
(D) Dotted edges are inhibited in the source node. The inhibitor here is CI, meaning a lytic attack requires any CI to become non-physiological by outside means (here: host-based proteolysis), see Figure~\ref{fig:abintra}:2/\{a,b\}/ vs.\ 2/c/.
(E) Reflexive edges indicate active self-regulation (of lysogeny), see Figure~\ref{fig:abintra}:2/a/.
(F) Boxes indicate `synchro-sustainability' (of lysogeny), see Figure~\ref{fig:abintra}:2/a/.

\item[Figure~\ref{fig:phasespace}:] Phenotypes exhibited by varying CI's intrinsic affinities for O\textsubscript{R}\{1,2,3\}, including the possibility that higher-than-wild-type concentration is needed for physiological binding. The origin in the  figure is for binding at low concentration (high affinity), with also medium, high, and extra-high concentrations shown. The wild type is indicated with bold border (low, high, high nominal concentrations). Green (upper left) is \emph{temperate}, red (upper right) \emph{virulent}, and yellow (lower) \emph{clear}, see Section~\refSectCombinator{}. The mixed-color squares are hybrids that likely will be identified as the most destructive phenotype (most used color).

\end{description}

\newpage\section*{Supporting Material}
\begin{description}
\item[Instructional video:] ``Using the CEqEA tool: \emph{A \emph{synthetic} Genetic Switch}'',\newline at \texttt{http://ceqea.sourceforge.net/extras/instructionalPoL.mp4} --- with insert from St-Pierre and Endy \cite{StPierreEndy:PNAS105} (with permission).

\item[\texttt{coext.v} (included):] Proof scripts for the Coq Proof Assistant \cite{coq} to formally verify the logical meta-theory of our CEqEA tool \cite{CEqEA}, see Section~\refSectCoext{}.

\item[\texttt{lambdaAGS3-RI.mig} (included):] Our MIG/RI specification for CEqEA that captures the premises that the reasoning in Ptashne \cite{Ptashne:AGS3} proceeds from. We refer to it as the $\lambda$\textsuperscript{[AGS3]}-genotype, see Section~\refSectGenotype{}. 

\item[\texttt{lambdaAGS3-cert.txt} (included):] CEqEA's certification of the form-to-function compilation of \texttt{lambdaAGS3-RI.mig}/the $\lambda$\textsuperscript{[AGS3]}-genotype, see Section~\refSectPhysiology{}.

\item[\texttt{lambdaAGS3-phys.mig} (included):] A reconstituted MIG specification of the physiological influences coded for by the $\lambda$\textsuperscript{[AGS3]}-genotype: a minimal but non-mutable organism account --- extracted from \texttt{lambdaAGS3-cert.txt}.

\item[\texttt{lambdaVar-generic.mig} (included):] 1,728 variants of \texttt{lambda[AGS3]\_RI.mig}, see Section~\refSectCombinator{}.
\end{description}

\vfill\section*{Acknowledgements}
RV thanks Jittisak Senachak for coding prototypes of CEqEA's form-to-function compiler, Olivier Danvy for comments on the manuscript, and Olivier Danvy, Hiroakira Ono, 
Kiyoyuki Terakura, and Mun'de Vestergaard for discussions. The authors declare no competing interests and no funding.

\section*{Contributions}
RV conceived of and did the work. EP advised on visualization, adapted his ZGRViewer tool for the requirements of interactive CCP usage in CEqEA, and made Figures~\ref{fig:computing}--\ref{fig:phasespace} and the video with RV.


\enddocument